\documentclass[USenglish]{article}
\usepackage{fancyhdr}
\usepackage[utf8]{inputenc}
\usepackage[small]{article}
\usepackage{microtype}
\usepackage{caption}
\usepackage{subcaption}

\begin{document}
  \author[1]{S.~Rajashankar}
  \author[2]{N.~Ananthkrishnan}
  \author[3]{A.~Sharma}
  \author[4]{J.~Lee}
  \author[5]{H.J.~Namkoung} 
  \affil[1]{Yanxiki Tech, Pune, Maharashtra, 411001, India. Email:rajashankar@yanxiki.com, rajashankars@alum.iisc.ac.in}
  \affil[2]{Yanxiki Tech, Pune, Maharashtra, 411001, India. Email:akn.korea.19@gmail.com, akn@aero.iitb.ac.in}
  \affil[3]{Yanxiki Tech, Pune, Maharashtra, 411001, India. Email:anurag@yanxiki.com.}
  \affil[4]{Hyundai Rotem, Uiwang-si, Gyeonggi-do, South Korea. Email:green32jw@hyundai-rotem.co.kr}
  \affil[5]{Hyundai Rotem, Uiwang-si, Gyeonggi-do, South Korea. Email:namkoung@hyundai-rotem.co.kr}
  \title{Turbojet Module Sizing for Integration with Turbine-Based Combined Cycle Engine}
  \abstract{A turbine-based combined cycle (TBCC) vehicle is studied that relies on a scramjet engine for high-speed flight but requires a turbojet module to accelerate it to a high supersonic handover Mach number. The challenge is to scale a given turbojet engine (TJE) core (compressor, burner, turbine) to a particular value of the air mass flow rate such that the desired thrust at the handover point is achieved. To this end, a model for the engine core is integrated with a supersonic intake model that is designed to supply the required mass flow rate, and a nozzle model that is expected to deliver the desired thrust. Both the TJE intake and nozzle are constrained by the design choices made for the DMSJ module, and the TJE core is itself constrained by the volume available from the TBCC vehicle sizing for hypersonic flight. The TJE module is sized by scaling the engine core with matching intake and nozzle designs in an iterative manner until the process converges to a solution with acceptable thrust satisfying all the system constraints. The task turns out to be non-trivial due to the scarcity of steady operating points for the engine core at high speeds, due to possible mismatch between the mass flow rate demanded by the compressor and that delivered by the supersonic intake, and due to the difficulty in adapting a DMSJ-style single-expansion ramp nozzle (SERN) to adequately expand the turbojet exhaust flow.}

\maketitle
\thispagestyle{empty}
\pagestyle{fancy}
\fancyhf{}
\fancyfoot[C]{\small \thepage} 
\section{Introduction} 
There has been sustained interest in recent decades in hypersonic cruise vehicles for both military and civilian applications (Bradley et al.~\cite{bradley2002revolutionary}, Kondo et al.~\cite{kondo1995overview}, Mamplata and Tang~\cite{mamplata2009technical}, Marshall et al.~\cite{marshall2004critical}, Moses et al.~\cite{moses2004nasa}, and Redifer et al.~\cite{redifer2007hyper}). However, a commercial hypersonic airliner needs to take off and land horizontally on a runway, which requires a propulsion system that can power the vehicle from take-off all the way to Mach 5 to 6, or higher. Ramjets and scramjets are the engines of choice for high supersonic flight above Mach 3, but they cannot operate at lower speeds. Hence, an alternative propulsion system is necessary to carry the vehicle from take-off to about Mach 2.5 to 3, and again bring it down to land. The typical choices are either a turbojet engine or a rocket, and accordingly these propulsion systems are called Turbine-based Combined Cycle (TBCC) or Rocket-based Combined Cycle (RBCC), respectively (Musielak~\cite{musielak2022scramjet}). Of these, TBCC engines are a more reasonable choice for commercial applications where both take-off and landing are required, and the engines are expected to be re-used.

Turbojet engines for high-speed flight up to Mach 3 have been used before in notable examples such as the J58 engine for the SR-71 and the Tumansky R-15 for the MiG-25, but neither of them truly qualifies as a TBCC engine. In the case of the J58, it has an afterburner, sometimes referred to as a ramjet, working in tandem with the turbojet core. However, this so-called ramjet does not operate independently of the turbojet. In a true TBCC engine, there are three distinct modes of operation: pure turbojet mode up to roughly Mach 3, after which the turbojet engine is shut down, followed by a ramjet mode up to approximately Mach 5, and then a scramjet mode for hypersonic flight. Several attempts are underway to design and test a TBCC engine with a dual-mode scramjet module (DMSJ) (McDaniel et al.~\cite{mcdaniel2009us}, Moses et al.~\cite{moses2004nasa}, and Redifer et al.~\cite{redifer2007hyper}), but to the best of our knowledge, no such engine is currently operational.

Since the mission of the hypersonic vehicle is long-range cruise, the design Mach number and altitude are typically around Mach 6 and 25-35 km. The vehicle is therefore sized for sustained cruise under these flight conditions. This means that the turbojet module of a TBCC system must accommodate the constraints that come from the design conditions at cruise. The most prominent of these are the following restrictions. Firstly, the geometric size of the vehicle is decided by lift and drag requirements at hypersonic cruise flight, so the turbojet must fit within the available volume. Two options are normally considered: a front-turbojet, back-scramjet tandem arrangement, or a top-turbojet, bottom-scramjet over-under arrangement (Chen et al.~\cite{chen2019research}, and McDaniel et al.~\cite{mcdaniel2009us}). Secondly, the engine intake, which is integrated with the vehicle forebody, is shaped for optimum performance at hypersonic cruise. The turbojet module needs to make best use of the available forebody ramp geometry, incorporating any variable-geometry features where unavoidable. Similarly, the turbojet module should make best use of the airframe-integrated single-expansion ramp nozzle (SERN) which is designed for hypersonic cruise conditions. Thirdly, the throat area of the scramjet channel will be set by the requirement of the starting condition, as given by the Kantrowitz criterion (Flock and G{\"u}lhan~\cite{flock2019modified}). This occurs at the lowest operating Mach number for the scramjet engine (in ram mode), usually around Mach 3. This presents a constraint when designing the intake duct for the turbojet channel. Finally, and perhaps most importantly, the turbojet engine should produce thrust close to the value of thrust from the ram mode of the dual-mode scramjet at the transition point, typically Mach 3, so that the handover from turbo to ram mode is smooth. Hence, sizing the turbojet module in a TBCC engine is not a straightforward matter.

In this work, we consider a dual-mode scramjet engine that has been sized for cruise flight at Mach 6, 27 km altitude with a starting Mach number of 3. Thus, the geometric features of the vehicle and the design of the scramjet channel are available. We assume an over-under configuration for the TBCC system. The problem then becomes one of sizing the turbojet module such that it fits in the available volume, produces the required thrust at the transition point from turbojet to ramjet mode (namely, at Mach 3), and can accommodate an air intake that is reasonably efficient and is able to take in the required air mass flow rate in order to produce this thrust. As a starting point, we take up the J85 turbojet engine for which data is available in the literature (Seldner et al.~\cite{Seldner1972}). It has also been incorporated in the early stages of the Hermeus Quarterhorse TBCC system development\footnote{https://www.hermeus.com/quarterhorse}. The engine modeling and analysis are carried out using T-MATS (Toolbox for the Modeling and Analysis of Thermodynamic Systems), an open-source package that runs on MATLAB\textsuperscript{\textregistered} (Chapman et al.~\cite{chapman2014a}). First of all, the J85 data is modeled and simulated in T-MATS and validated with the literature (Seldner et al.~\cite{Seldner1972}). It is then tested for supersonic operation at Mach 2.5 and 3.0. Next, we demonstrate how the turbojet engine model may be scaled in T-MATS for any given value of air mass flow rate. Having prepared the model, we can carry out the sizing of the turbojet engine (TJE) to match the given DMSJ module of the TBCC system, which is the main focus of this work. Specifically, once we know the thrust required from the TJE at the turbojet-ramjet transition point, we need to find the air mass flow rate corresponding to this thrust, and the available J85 turbojet engine model is then scaled to this air mass flow rate. The turbojet intake, both the supersonic and subsonic diffuser sections, is then designed within the constraints placed by the DMSJ module, to provide the desired total pressure recovery and air mass flow rate to the TJE module. Following this, a Single-Expansion Ramp Nozzle (SERN) is designed with a thrust calculation sub-routine to verify that the required thrust is indeed achieved. The liquid fuel injected in the turbojet is JP-10, same as that used for the DMSJ, with a lower heating value (LHV) of 43 MJ/kg (Tao et al.~\cite{tao2018physics}). The supersonic intake and the SERN nozzle are simulated using an approximate inviscid solution methodology called Reduced-Order Aerodynamic Modeling (ROAM), which is based on the work in Dalle et al.~\cite{2010Dalle-1}. Once the methodology is established, the turbojet core, intake, and nozzle are integrated for a given air mass flow rate. The intake is sized for cowl-lip height ($h_c$), throat height ($h_{2t}$), and intake diffuser exit diameter ($d_3$) to provide the selected air mass flow rate to the turbojet core. The turbojet core is scaled (scale factors for performance maps, and nozzle throat diameter ($d_{6t}$)) to the flow delivered by the intake. This is followed by the nozzle sizing where the main parameter is the diffuser exit height ($h_{6d}$). This process is repeated iteratively until convergence when the thrust calculated from the sizing process matches the ram mode handover thrust (within an error margin). Finally, a tip-to-tail analysis of the scaled TJE engine is carried out both under design and off-design conditions, and it is shown that the engine performance is as desired. Crucially, finding a steady state on the TJE operating line at high Mach numbers and altitudes can be a challenge. At higher altitudes, the stall line and the operating line typically approach each other, increasing the risk of stalling during acceleration maneuvers. Besides, the sensible heat addition in the TJE combustor at higher Mach numbers may be limited, which means that the range of operating points could be quite narrow. Thus, very careful matching of the TJE components is necessary to ensure TJE acceleration to Mach 3 and the subsequent handover to the ram mode of the DMSJ module.

Typical analyses in the past have involved selecting models for the turbojet and the high-speed modules, which are then used to evaluate the overall TBCC engine performance (Trefny and Benson~\cite{trefny1995integration}, Ma et al.~\cite{MA2018141}, and Yu and Guo~\cite{YU9188480}). There are some studies on the design and performance analysis of the intake (Sanders and Weir~\cite{Sander2008}, and Yuan et al.~\cite{Yuan2020}) and the nozzle (Lv et al.~\cite{lv2022recent}) for turbojet engine integrated with a high-speed intake/ nozzle. This paper is perhaps the first instance, to the best of our knowledge, where the modeling, scaling, integration, and performance analysis of the turbojet engine in a TBCC system has been the focus of attention. In the following sections, details of the DMSJ module and the problem definition are stated in Sec.~\ref{sec:Prob_defn}. This is followed by modeling and validation of the baseline turbojet engine in Sec. \ref{sec:TJE_core}, and its high-speed extension and scaling are discussed therein. TJE nozzle modeling is presented in Sec.~\ref{sec:Nozzle}, where application of the ROAM methodology to the nozzle flow is also presented. This is followed by the modeling of the the intake in Sec.~\ref{sec:Intake}. The sizing of the TJE is then carried out in an iterative manner with the inlet and nozzle integrated with the turbojet core in Sec.~\ref{sec:TJE_sizing}. Finally, the performance of this scaled TJE engine is analyzed under design and off-design conditions.

\section{Problem Formulation} \label{sec:Prob_defn}
In this section, we shall first present the DMSJ module that has already been designed. The DMSJ module of the TBCC system is sized for optimal performance at cruise conditions of Mach 6 at an altitude of 27 km. This imposes constraints on the sizing, performance, and efficiency of the turbojet module of the system.  At the same time, the TJE module is required to deliver a thrust very close to the thrust of the DMSJ ram mode at the handover Mach number of 3. 

\begin{figure}[tb]
    \centering
    \includegraphics[width=1\linewidth]{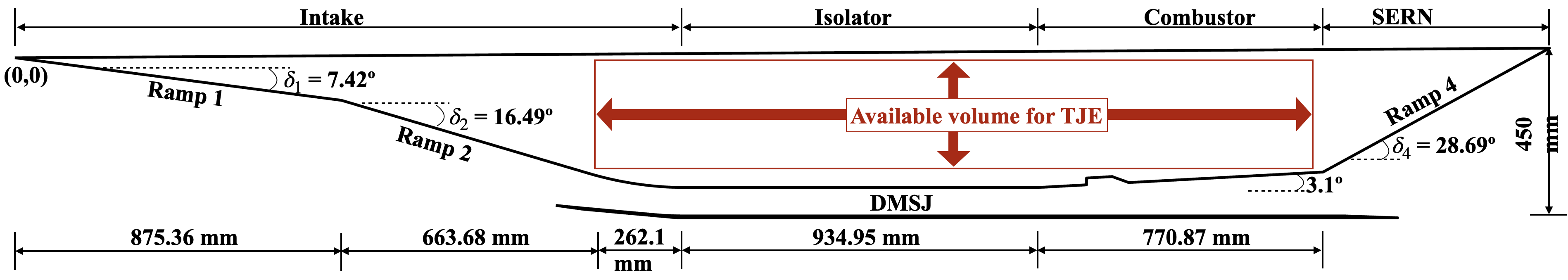}
    \caption{DMSJ configuration with components and dimensions marked.}
    \label{fig:DMSJ_schematic}
\end{figure}

The schematic of the designed DMSJ vehicle is illustrated in Fig.~\ref{fig:DMSJ_schematic}.
The DMSJ configuration comprises of an airframe-integrated double ramp intake, isolator, combustor, and an airframe-integrated single-expansion ramp nozzle. As the name suggests, the dual-mode scramjet operates in two modes: Ram mode at low supersonic speeds (likely from Mach 3 to Mach 5), with combustion occurring at subsonic speeds, and supersonic mode (likely beyond Mach 5), where the combustor operates at supersonic speeds. As depicted in Fig.~\ref{fig:DMSJ_schematic}, the intake is a double-ramp supersonic intake designed to efficiently facilitate the cruise operation of the hypersonic vehicle at Mach 6 and 27 km altitude. The first ramp angle is $\delta_1=7.4$ deg, followed by a second ramp of angle $\delta_2=16.5$ deg (incremental ramp angle of $9.1$ deg). These ramp angles are optimized using the Oswatitsch criterion (Oswatitsch~\cite{oswatitsch1980pressure}, and Ara{\'u}jo et al.~\cite{araujo2021optimization}) for Mach 6, which ensures that the total pressure recovery across each of the ramp shocks is the same. Following the intake ramps is a constant-area isolator, whose length is dictated by the need to ensure that the terminal normal shock in ram mode is held within the isolator under all conditions. A circular arc is incorporated to ensure a smooth transition from the second ramp to the isolator. The isolator is succeeded by the combustor, featuring a diverging upper wall angle of $3.1$ deg. Subsequently, the single expansion ramp nozzle follows the combustor with expansion angle $\delta_4=28.7$ deg.

The transition from turbojet to ram mode of the DMSJ is planned for Mach 3 at altitude 18.1 km, which corresponds to a dynamic pressure of 46.55~kPa. Assuming these conditions are held constant, with an angle of attack of zero, the ram mode thrust is known from the DMSJ analysis to be 1568~N. Hence, the design requirement for the TJE module becomes that of achieving this thrust (with an allowable error margin) so that the vehicle transitions smoothly from TJE to DMSJ mode. Additionally, the DMSJ design imposes the following constraints on the TJE configuration:
\begin{itemize}
    \item (C1) The TJE must be sized to fit within the available volume in the airframe above the DMSJ engine, marked by a rectangular box in Fig.~\ref{fig:DMSJ_schematic}. Since the mass flow rate through the core scales as the square of the engine diameter, and the thrust is a function of the mass flow rate, this effectively limits the thrust output of the TJE.
    \item (C2) The TJE must have a steady operating point at Mach 3, 18.1 km altitude flight. As pointed out earlier, this can prove to be a bit of a challenge. Typically, at high supersonic Mach numbers, one may expect steady states over only a very narrow range of fuel flow rates (equivalence ratios).
    \item (C3) The TJE intake ramp angles are constrained by the values selected for optimal DMSJ operation at Mach 6. Invariably, those angles are too small for the optimal TJE intake at Mach 3 and below. Consequently, the compression process due to the intake is less than desired, and the total pressure recovery is also not optimal. Additional compression must take place in the supersonic diffuser duct, which means a longer duct length is needed than otherwise.
    \item (C4) The TJE shares the nozzle exit ramp with the DMSJ,  the ramp angle being optimized for DMSJ thrust. In case of the DMSJ in scram mode, the nozzle flow typically expands from approximately Mach 2 to an exit value around Mach 5, whereas the same geometry has to expand the TJE flow from sonic to a value just over Mach 2. For the TJE intake as well as the nozzle, the respective splitter plates can be properly positioned to recover as best performance as possible.  
\end{itemize}
In summary, due to the various constraints, and given the Mach number and altitude of operation, it is quite likely that the TJE may not have a steady operating point at all, and even if it does, the intake may not be able to supply the required air mass flow rate, and even if that is possible, the nozzle expansion may not yield the desired thrust. Thus, the TJE design for integration into a TBCC engine is not a trivial affair.

In this work, we approach the problem in three steps. Step 1 is to ensure that the TJE does indeed have one or more steady operating points at the given Mach, Altitude conditions within the allowable range of fuel flow rates. We shall check the performance of the turbojet core for Mach numbers 2.5 and 3 since they are the most challenging. Also Mach 2.5-3.0 is the typical range where the transition from turbojet to ram mode takes place. After that, we check whether the TJE can be scaled to fit within the given volume in the TBCC engine. The scaling is done with respect to the mass flow rate. Since the thrust calculation comes from the nozzle module, at this step we still do not know whether the scaled TJE provides the desired handover thrust. That is, the final sizing cannot be done until the nozzle module is completed and the thrust calculated, which in turn requires the values from the exit of the TJE core as input conditions. In Step 2, the nozzle calculations are completed and the thrust is evaluated for various mass flow rate values. At this stage, it is possible to map the desired handover thrust to a mass flow rate value through the TJE core. However, we do not yet know if the intake will allow this mass flow rate to enter the TJE core, hence the intake must also be integrated with the TJE core and nozzle before the final calculations can be made. Then Step 3 is to design the intake ramps and splitter, and the diffuser shape to provide the required mass flow rate with the corresponding total pressure recovery. After this intake design is integrated with the TJE core and the nozzle, some adjustments may still be needed in the final sizing. Once that is successful at the design condition, the final step is to evaluate the TJE performance for off-design cases. 

\section{Modeling the TJE Core} \label{sec:TJE_core}
\begin{figure}[tb]
    \centering
    \includegraphics[width=0.9\linewidth]{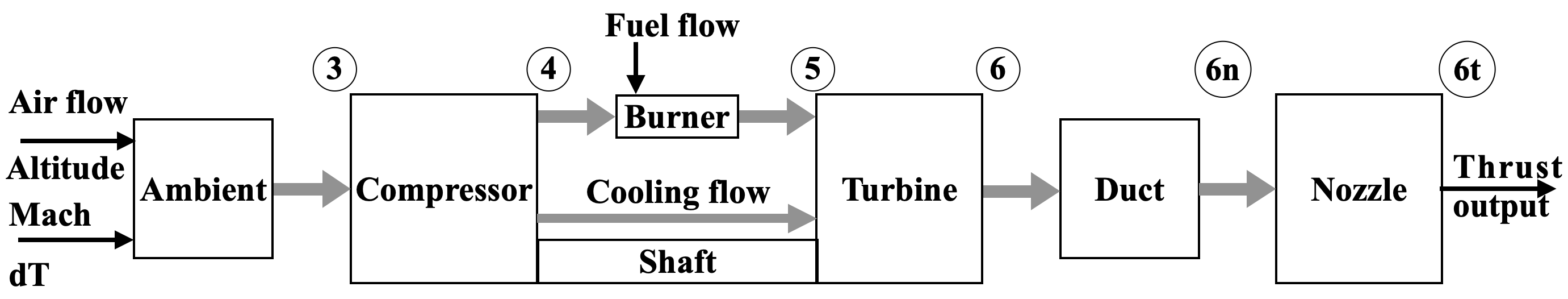}
    \caption{J85 turbojet engine architecture in T-MATS.}
    \label{fig:J85_Architecture}
\end{figure}

To model the core stages of the TJE, we select the J85 turbojet as our baseline engine. The J85 is a small, single-shaft turbojet engine, extensively used in military applications, with an eight-stage compressor, a two-stage turbine, and a variable-area nozzle. The maximum dry thrust of the engine at sea level is approximately 12~kN with a corresponding mass flow rate of 19.96~kg/s (Chapman et al.~\cite{Chapman2016}). Using the compressor and turbine maps for the J85 available in the literature (Kopasakis et al.~\cite{Kopasakis2008}, Seldner et al.~\cite{Seldner1972}), we propose to scale the engine model for our desired mass flow rate, Mach and altitude operation. The turbojet engine modeling is carried out by using the open-source Toolbox for the Modeling and Analysis of Thermodynamic Systems (T-MATS) developed by NASA Glenn Research Center (Chapman et al.~\cite{chapman2014a}). The block diagram of the J85 turbojet engine model in T-MATS is shown in Fig.~\ref{fig:J85_Architecture}.
The TJE core components are interconnected as shown in Fig.~\ref{fig:J85_Architecture}: intake (ambient), followed by the high-pressure compressor, then the burner (combustor), which is linked to the high-pressure turbine, a duct segment, and finally the exhaust nozzle. The high-pressure compressor and turbine are connected via a shaft. The various station numbers are marked in Fig.~\ref{fig:J85_Architecture}.

The modeling of turbojet engine components in T-MATS is discussed in detail in Chapman et al.~\cite{Chapman2016} and is not reproduced here. Briefly, the model equations follow a thermodynamic cycle approach based on the Brayton cycle, ensuring the conservation of energy and momentum throughout the system. Compression and expansion processes are treated as isentropic and adiabatic, while combustion occurs at constant pressure. Compressor and turbine performance are simulated using performance maps, which are empirical, nonlinear models. These maps can be generated using computational fluid dynamics, experimental data, or generic estimations. Utilizing pre-generated generic maps allows for scalability to different applications and offers a convenient method for modeling unknown turbomachinery. However, while scaled maps may provide satisfactory matching near the design point, significant errors can accumulate as the engine operates across its operational envelope. The entire system must be balanced for work between the compressor and turbine, and must, of course, satisfy the conservation of mass and energy. Turbomachinery is often characterized by two quantities, namely, flow coefficient ($\phi$) and work coefficient ($\Psi$), which are given by 
\begin{equation} \label{Eq:flow_coeff}
    \phi\ =\ \frac{C}{U}
\end{equation}
and
\begin{equation} \label{Eq:work_coeff}
    \Psi\ =\ \frac{\Delta H}{U^2}
\end{equation}
where, $C$ is the axial velocity to the rotor, $\Delta H$ is the change in stagnation enthalpy, and $U$ denotes rotor velocity. The work balance between the compressor and the turbine can be achieved by equating the work coefficients of these two components. This is performed internally in T-MATS where a steady-state Newton-Raphson solver is employed to achieve convergence of the system and determine the steady operating points on the compressor and turbine maps. Initially, component variables including mass flow rate, \textit{Rline} for the compressor, pressure ratio for the turbine, and shaft speed are estimated. Subsequently, these variables undergo adjustment during each iteration, facilitated by the monitoring of normalized flow rates from the compressor, turbine, and nozzle, as well as the rate of change of shaft speed. Iterations continue until the monitored variables reach the predetermined convergence criteria, set to the absolute value of 0.005 in this study.

\subsection{Validation of J85 T-MATS Model}
The J85 engine has previously been modeled in T-MATS and validated using experimental data from the literature by Chapman et al.~\cite{Chapman2016}. Generic performance maps for the compressor and turbine were utilized, and the performance characteristics of the J85 engine were studied, considering both sea-level and cruise conditions. For the current study, the architecture for the reference engine model is derived from the study by Chapman et al.~\cite{Chapman2016}, and the performance maps of the high-pressure compressor and high-pressure turbine for the J85 engine are sourced from Kopasakis et al.~\cite{Kopasakis2008} and Seldner et al.~\cite{Seldner1972}, respectively. The J85 high-pressure compressor performance map is as depicted in Fig.~\ref{fig:J85_PerformanceMap_Compressor}: 
\begin{figure}[tb]
    \centering
    \includegraphics[width=0.5\linewidth]{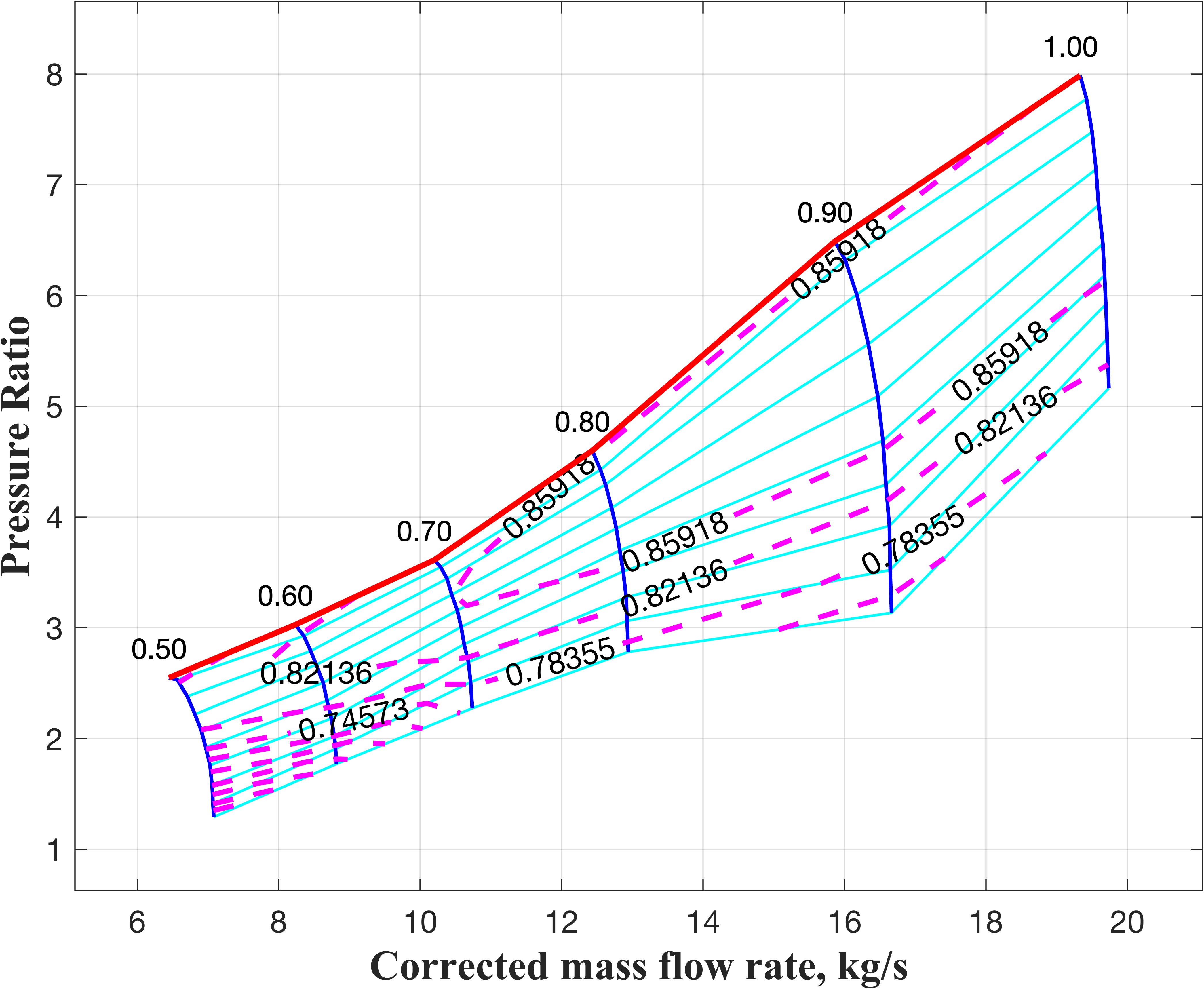}
    \caption{Performance map for baseline engine high-pressure compressor.}
    \label{fig:J85_PerformanceMap_Compressor}
\end{figure}
The lines for different corrected shaft speeds ($N_c$) from 0.5 to 1.0 are marked in blue. The corrected speed curves are normalized with respect to the design corrected speed. The locus of constant efficiency (\emph{Eff}) points are connected by dashed pink lines, while the surge line is marked in bold red. The {\it Rline} are shown in green, displaced from the surge line, with values starting at 1.0 at the surge line and incrementing with a step of 0.2 towards lower pressure ratio (PR). The high-pressure turbine performance map is similarly depicted in Fig.~\ref{fig:J85_PerformanceMap_Turbine}.
\begin{figure}[tb]
    \centering
    \includegraphics[width=0.5\linewidth]{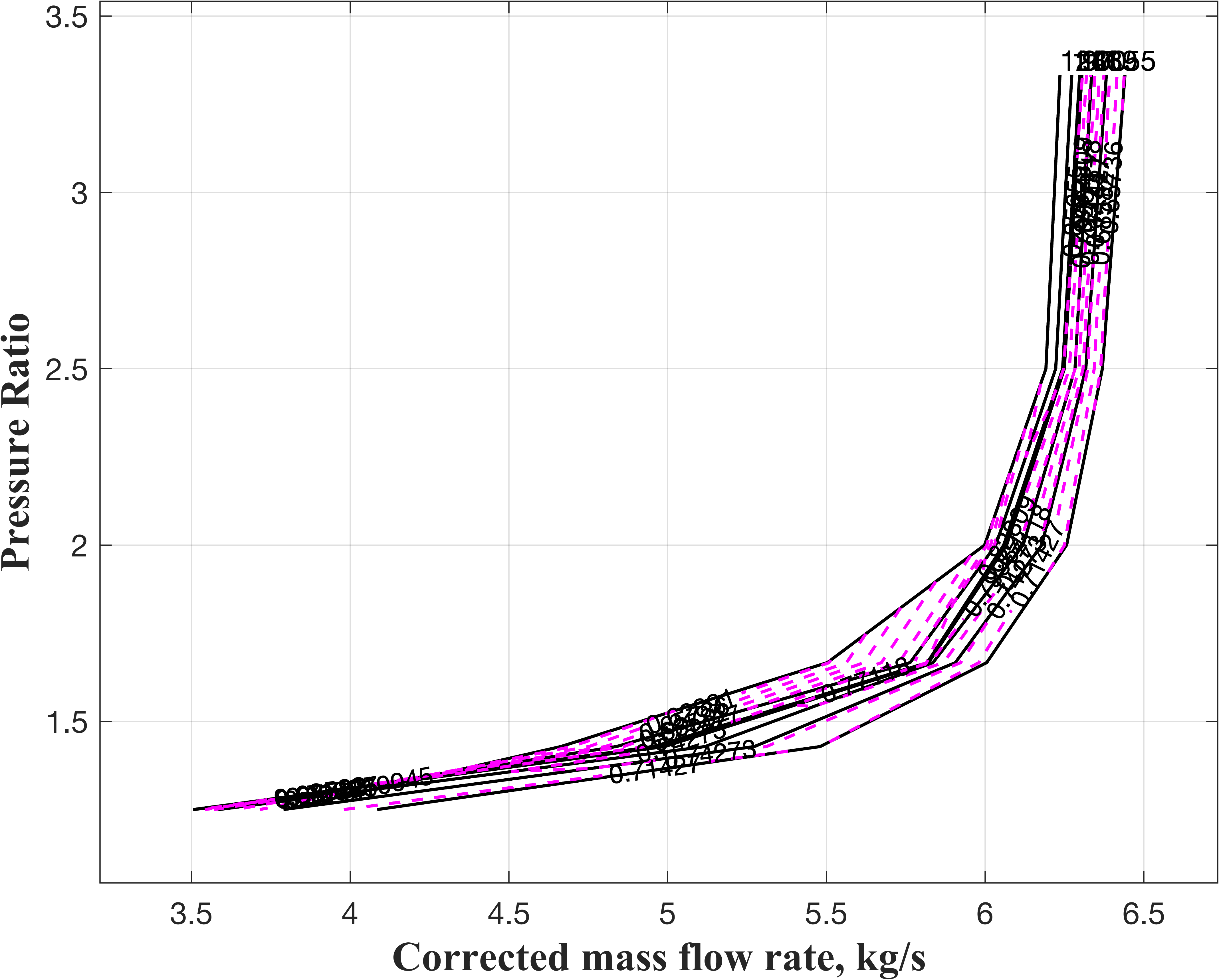}
    \caption{Performance map for baseline engine high-pressure turbine.}
    \label{fig:J85_PerformanceMap_Turbine}
\end{figure}

The test case for validation is taken from Seldner et al.~\cite{Seldner1972}, and is as noted in Table~\ref{tbl:J85_design_condition}. This corresponds to Mach number zero at sea level. The data point for the 100\% corrected shaft speed case is considered as the design point. 
\begin{table}[b!] 
    \centering
    \caption{Baseline turbojet engine parameters at original design point}
    \label{tbl:J85_design_condition}
    \begin{tabular}{p{3.5cm}l|p{3.5cm}l}
    \multicolumn{2}{c|}{\textbf{Compressor}} & \multicolumn{2}{c}{\textbf{Turbine}} \\
    \hline
    $Mass\ flow,\ kg/s$ &	$19.913$ & $Map\ design\ N_c$ &$	100$\\
    \it{Map\ design\ efficiency} &	$0.834$ & \it{Map\ design\ efficiency} & 	$0.862$\\
    $Map\ design\ N_c$ &	$1 $& $Map\ design\ PR$ &	$2.653$\\
    $Map\ design\ PR $&	$6.946$ & $Design\ N,\ rpm$ &	$16540$\\
    $Map\ design\ Rline$ &	$1.7201$ & $TIT,\ K$ &	$1225.6$ \\
    $Bleed\ air\ in\ fractions$	& $0.0182\ to\ turbine$  & \multicolumn{2}{c}{}  \\
      &  $0.0348\ out$  &  \multicolumn{2}{c}{}  \\
    \hline
    \multicolumn{2}{c|}{\textbf{Burner}} & \multicolumn{2}{c}{\textbf{Nozzle}}\\
    \hline
    $LHV,\ MJ/kg$ &	$42.8 $ &$ Nozzle\ mass\ flow,\ kg/s$ &	$19.6$\\
    \it{Efficiency} &	$0.98$  &  \multicolumn{2}{c}{}\\
    $Pressure\ loss\ \%$ &	$7.21$  &  \multicolumn{2}{c}{}\\
    $Fuel\ mass\ flow, kg/s$ &	$0.369$ &  \multicolumn{2}{c}{}\\
    \end{tabular}
\end{table}
The engine model in Fig.~\ref{fig:J85_Architecture} is run with the data in Table~\ref{tbl:J85_design_condition}, and 
the performance variables are calculated at the design point by disabling the {\it iDesign} option. A comparison of the results from this validation run with the data from Seldner et al.~\cite{Seldner1972} and Chapman et al.~\cite{Chapman2016} is presented in Table~\ref{tab:validationofTmats}, where $P$ and $T$ represent the pressure and temperature, respectively. The match can be seen to be excellent with the error being of the order of 1\% or lower.
\begin{table}[h]
    \centering    
    \caption{Validation of T-MATS for baseline turbojet engine}
    \label{tab:validationofTmats}
    \begin{tabular}{p{4.25cm}p{2cm}p{2cm}p{2cm}}
      \textbf{Parameter} &	\textbf{Value from Ref.~\cite{Seldner1972, Chapman2016}}	& \textbf{Value from present T-MATS run}  & \textbf{\% Error}\\
      \hline
    $Design\ N_c\ at\ 100\%,\ rpm$	&$16540$	&$16537.278$	&$0.0165$\\
   $ Compressor\ out,\ T\ (K)$&	$543.9$&	$540.3$&	$0.6639$\\
    $Compressor\ out,\ P\ (Pa)$&	$703816.8$&	$703265$&	$0.0784$\\
    $Mass\ flow\ without\ bleed\ (kg/s)$ &	$19.913$ &	$19.919$ &	$-0.03189$\\
    $Burner\ out,\ T\ (K)$&	$1225.6$ &	$1225.6$&	$0$\\
    $Burner\ out,\ P\ (Pa)$&	$653071.41$&	$652795.62$&	$0.04222$\\
    $Burner\ mass\ flow\ (kg/s)$&	$19.232$&	$19.232$&	$0$\\
   $Turbine\ out,\ T\ (K)$&	$991.7$&	$996.1$&	$-0.4482$\\
    $Turbine\ out,\ P\ (Pa)$&	$246142.8$&	$246694.42$&	$-0.2241$\\
    $Nozzle\ mass\ flow\ (kg/s)$&	$19.6$&	$19.6$&$	0$\\
    $Thrust\ (kN)$&	$12.1$ &	$12.26$&	$-1.3235$\\
    \end{tabular}
\end{table}

\subsection{Extension to High-Speed Operation}
The J85 baseline model is now run in T-MATS to determine the steady operating points at Mach 2.5, Altitude 15.8 km, and at Mach 3.0, Altitude 18.1 km. Since the operating points will presumably occur over a range of air mass flow rates, and the intake has not yet been designed, a stop-gap measure is necessary to provide the ambient/ intake values at the entry to the compressor face. Therefore, an alternative high-speed intake operating in critical mode is attached upstream of the TJE core. It must be emphasized that this is not related to the actual intake in Fig.~\ref{fig:DMSJ_schematic}.

\begin{figure}[b]
    \centering
    \includegraphics[width=0.8\linewidth]{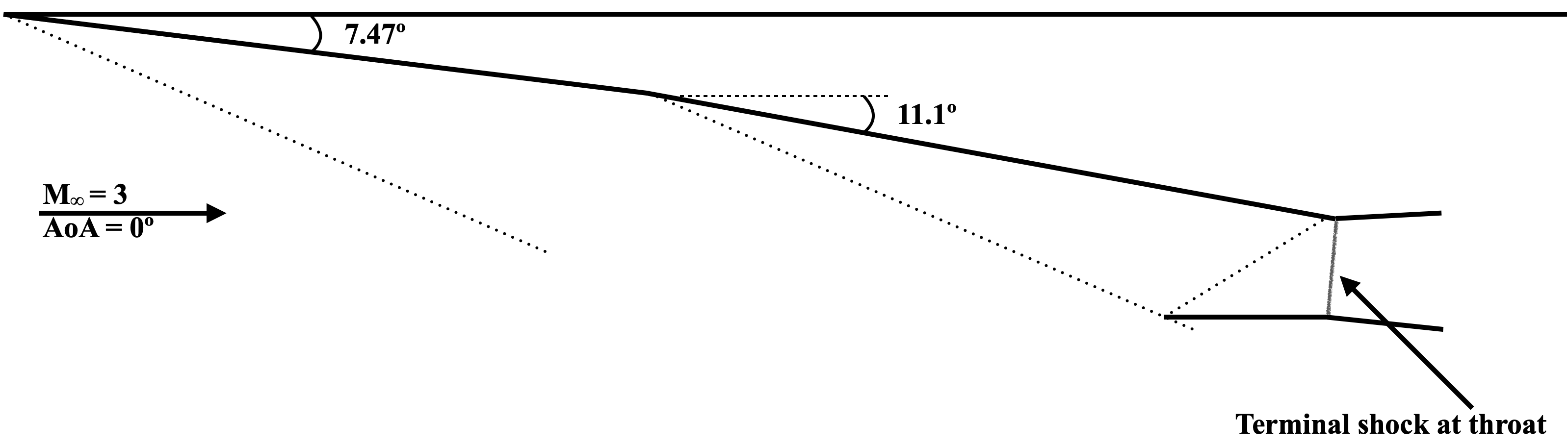}
    \caption{Supersonic intake geometry used for testing baseline engine model at high speeds.}
    \label{fig:supersonic_intake}
\end{figure}

For the case of operation at Mach 3, Altitude 18.1 km, the intake ramp angles and shock system are as shown in Fig.~\ref{fig:supersonic_intake}.
The first and second ramp angles are $7.5$ deg and $11.1$ deg, respectively, with the second ramp shock sitting on the cowl lip. An oblique shock is located at the cowl lip directed into the duct, and is followed by a terminal normal shock, as indicated in Fig.~\ref{fig:supersonic_intake}. The total pressure loss across this intake shock system is evaluated, as given in Table~\ref{tab:amb_terminalproperties}, though no additional viscous losses are considered. In a similar manner, the intake ramp and shock system are artificially created for the case of operation at Mach 2.5, Altitude 15.8 km. The intake flow properties calculated in this manner, downstream of the terminal shock, are given in Table~\ref{tab:amb_terminalproperties} for the two cases of free stream Mach number, 2.5 and 3.0, at their corresponding altitudes (subscripts $\infty$ and $0$ represent free-stream and total conditions, respectively).
\begin{table}[b]
    \centering
    \caption{Calculated intake properties downstream of  terminal shock used to test baseline engine model at high speeds}
    \label{tab:amb_terminalproperties}
    \begin{tabular}{lll}
        \textbf{Properties}	& \textbf{M$_\infty$ = 3,\ Alt=18.1~km} &	\textbf{M$_\infty$ = 2.5,\ Alt=15.8~km}\\
        \hline
        $P,\ Pa$ &	$147355$ &	$115345$\\
        $T,\ K$ &	$569.62$ &	$449.46$\\
        $P/P_\infty$ &	$19.947$ &	$10.864$\\
        $T/T_\infty$ &	$2.626 $&	$2.072$\\
        $P_0/P_{0\infty}$ &	$0.679 $&	$0.848$\\
        $M$ &	$0.575$ &	$0.654$\\
    \end{tabular}
\end{table}
These are then used to evaluate the flow properties at the compressor face, which serves as input to the high-pressure compressor block in T-MATS.
 
\begin{figure}[tb]
    \centering
    \includegraphics[width=0.5\linewidth]{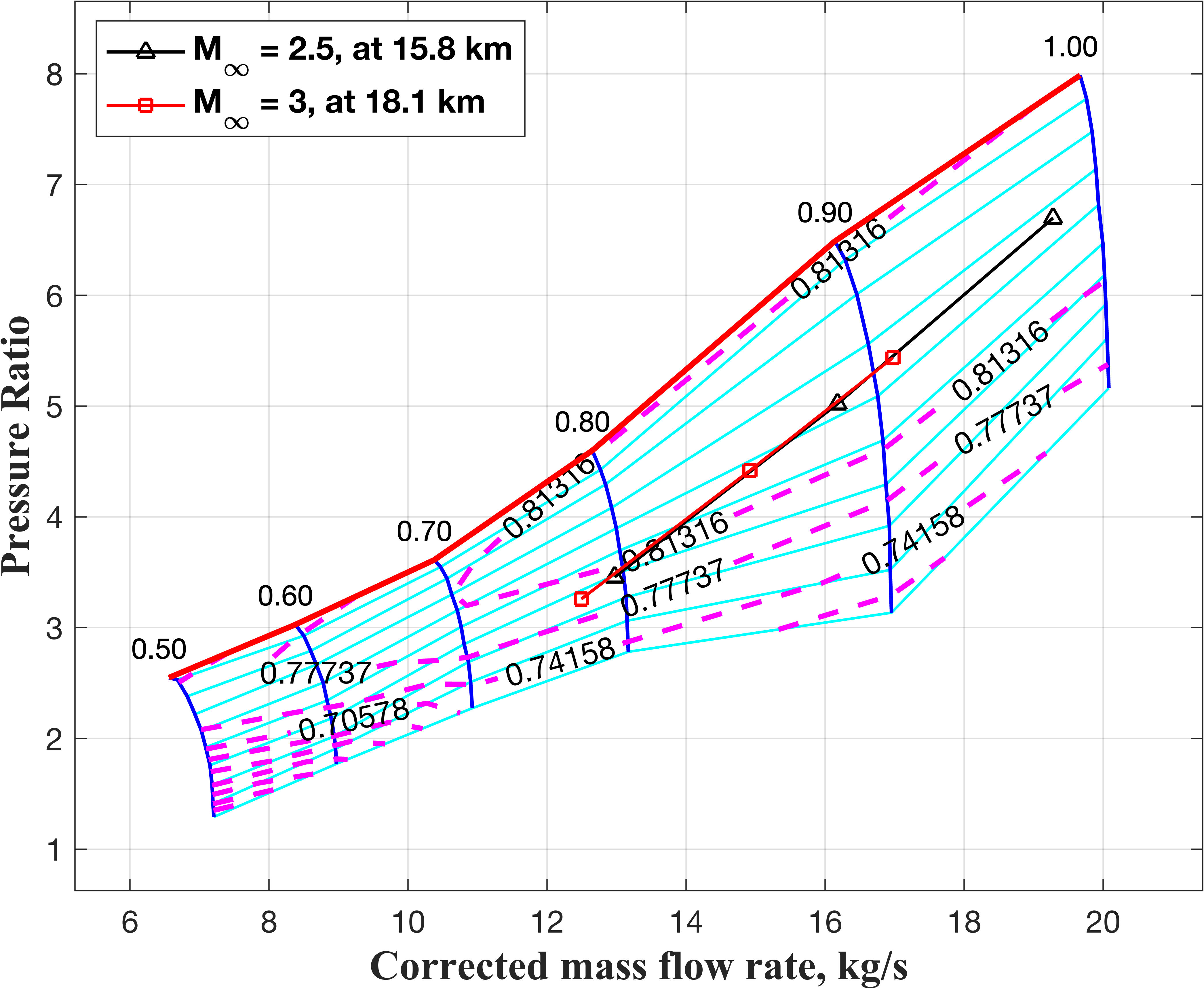}
    \caption{Operating line with steady states for the baseline engine in high-speed operation.}
    \label{fig:J85_offdesign_compressor_map}
\end{figure}

A critical constraint in high-speed operation is the limiting value of the Turbine Inlet Temperature (TIT). 
This constraint severely limits the amount of fuel that can be added in the burner making it difficult to obtain steady operating points where the work coefficient in Eq.~\ref{Eq:work_coeff} may be matched between the compressor and the turbine. For this calculation, the turbine inlet temperature is limited to 2100 K for both Mach 2.5 and Mach 3.0 cases. The operating lines for these two cases are constructed by varying the fuel mass flow rate and allowing T-MATS to determine the steady operating points. The operating points thus obtained are plotted on the compressor performance map shown in Fig.~\ref{fig:J85_offdesign_compressor_map}. 
Three operating points each are plotted for the Mach 2.5 and Mach 3.0 cases. The operating line for the two Mach cases is virtually identical. The extreme points on the operating line are limited by maximum turbine inlet temperature (TIT) and the convergence criterion at maximum corrected mass flow rate ($W_c$), and by the T-MATS convergence criterion at lower $W_c$. Failure of the convergence criterion implies that the compressor and turbine work matching could not be achieved.

\subsection{Scaling with Mass Flow Rate}
The baseline engine with the design nozzle mass flow rate of $19.6$~kg/s needs to be scaled down to, i) fit within the space available in the TBCC configuration, and ii) to match the available ramjet thrust at the handover Mach number of 3. In the following, we focus on scaling the TJE for a desired mass flow rate. Since the desired mass flow rate for the required thrust is not yet available at this stage, we prepare the scaling process for an arbitrary nozzle mass flow rate ($W$) of $6.9$~kg/s at sea level, Mach 0, with a mass scaling ratio of $0.352$. Later, when the thrust calculations are in place, an iterative procedure is followed to obtain the desired handover thrust, and the sizing calculations are carried out to correctly match all the TJE components (intake, engine core, and nozzle) which yields the correct value of scaled mass flow rate and the corresponding thrust.

The scaling of the turbojet engine in T-MATS is performed by adjusting the design mass flow through the nozzle and obtaining the corresponding scale factors for the performance maps. However, some corrections may be necessary to account for the bleed air and fuel mass addition. Also, as the air mass flow rate is scaled, the added fuel mass flow rate will also have to be modified. The scale factors at design conditions are obtained by enabling the {\it iDesign} option in T-MATS and by adjusting the fuel flow to get the same value of design TIT. The design conditions used for obtaining map scale factors by enabling {\it iDesign} are those already listed in Table~\ref{tbl:J85_design_condition}. The scale factors obtained in this manner for the compressor and turbine maps are listed in Table~\ref{tab:map_scalars}.
\begin{table}[b]
    \centering
    \caption{Scale factors for baseline engine performance maps for $W/W_c=0.352$}
    \label{tab:map_scalars}
    \begin{tabular}{p{2cm}llll}
        Scale Factor & \textbf{Compressor} & \textbf{Turbine} & \textbf{Compressor} & \textbf{Turbine}\\ \hline
        $N_c$ & ${N_{c\_perf}}/{N_{c\_map}}$ & $\frac{N_{c\_perf}/\sqrt{T_{std\_SL}}}{N_{c\_map}}$ & $16540$ & $3.522$ \\ 
        $W_c$ & \multicolumn{2}{c}{${W_{c\_perf}}/{W_{c\_map}}$}  & $1.017$ & $1.525$\\ 
        $PR$ & \multicolumn{2}{c}{${(PR_{perf}-1)}/{(PR_{map}-1)}$}  & $1$ & $0.996$\\ 
        \it{Eff} & \multicolumn{2}{c}{$({\it{Eff}_{perf}-1)}/{(\it{Eff}_{map}-1)}$}  & $0.946$ & $1$\\ 
    \end{tabular}
\end{table}
It may be noted that the compressor map is scaled such that the scale factor for Pressure Ratio is held at 1.0, whereas the turbine map is scaled with respect to the efficiency. If properly done, then the dimensions, speeds, air and fuel mass flow rates should be scaled, while the flow properties such as pressure and temperature should remain practically the same as those for the original sizing. The mass flow rate, static pressure and temperature at various stations of the original TJE and its scaled variant are reported in Table \ref{tab:flow_7kgs}. 
\begin{table}[tb]
    \centering
    \caption{Comparison of flow properties at different engine stations for design nozzle mass flow of 19.6~kg/s and scaled nozzle mass flow rate of 6.9~kg/s}
    \label{tab:flow_7kgs}
    \begin{tabular}{p{1.8cm}|p{1.1cm}p{0.8cm}p{1.2cm}|p{1.1cm}p{0.8cm}p{1.2cm}}
         & \multicolumn{3}{c|}{\textbf{W = 19.6 kg/s}} & \multicolumn{3}{c}{\textbf{W = 6.9 kg/s}}\\
         \hline
        	            & $W_c\ (kg/s)$	& $T\ (K)$	& $P\ (Pa)$	& $W_c\ (kg/s)$	& $T\ (K)$	& $P\ (Pa)$	\\
         \hline
        Compressor entry	& $19.92$	& $288.15$	& $101325.35$	& $7.0$	& $288.15$	& $101325.35$	\\
        Compressor Exit	    & $18.86$	& $540.23$	& $703513.46$	& $6.63$	& $540.35$	& $703740.98$ \\
        Burner Exit	        & $19.23$	& $1225.34$	& $652788.72$	& $6.76$	& $1225.64$	& $653002.46$	 \\
        Turbine Exit	    & $19.6$	& $995.96$	& $246715.1$	& $6.9$	& $996.112$	& $246756.47$	 \\
        Nozzle throat area ($m^2$)	& \multicolumn{3}{c|}{$0.07023$}	& \multicolumn{3}{c}{$0.02467$}	\\
         Engine diameter ($m$) & \multicolumn{3}{c|}{$450.0$}	& \multicolumn{3}{c}{$266.8$}	\\ 
    \end{tabular}
\end{table}
The temperature and pressure may be seen to be virtually identical between the original and the scaled mass flow rates, whereas the nozzle area has been downsized for the lower mass flow rate with the same ratio of $0.352$. The ratio of the scaled engine diameter may be checked to be precisely the square root of the mass flow scaling ratio of $0.352$. This procedure can be repeatedly applied for various values of the mass flow rate and mapped to the corresponding thrust values in an iterative procedure once the nozzle modeling is complete.

\section{Nozzle Modeling and Thrust Calculation} \label{sec:Nozzle}
\begin{figure}[tb]
    \centering
    \includegraphics[width=0.8\linewidth]{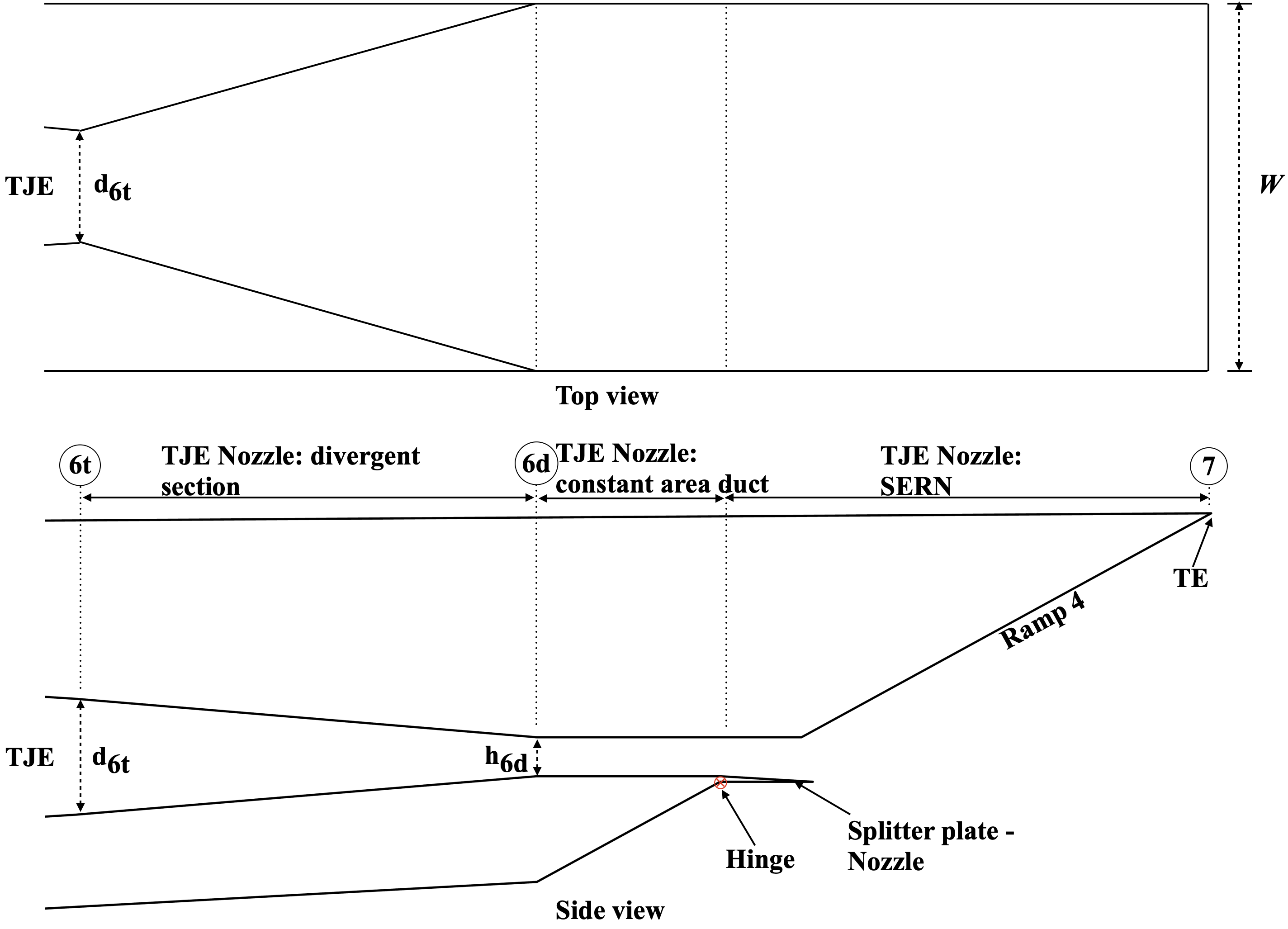}
    \caption{Schematic of the TJE nozzle segments downstream of the nozzle throat.}
    \label{fig:Nozzle_modeling}
\end{figure}

The TJE nozzle consists of four segments: a circular convergent duct up to the throat, which is modeled as part of the turbojet model in T-MATS, following the turbine; a divergent circular to rectangular transition segment; a constant-area duct with rectangular cross-section; and finally the single-expansion ramp (SERN). The latter three segments are as shown in Fig.~\ref{fig:Nozzle_modeling}, where $w$, $d_{6t}$ and $h_{6d}$ represent the width of the engine, the nozzle throat diameter, and the exit height of the divergent duct, respectively.
The divergent section is sized for an exit Mach number of 1.4, which gives an area ratio of 1.1266 considering isentropic flow with hot gas properties. The upper wall of the SERN is a single ramp of angle $\delta_4 = 28.7$ deg during DMSJ operation, which then blocks the exit of the TJE nozzle. For TJE operation, a section of this ramp is designed to be a short splitter plate that can be rotated about a hinge, as marked in Fig.~\ref{fig:Nozzle_modeling}, which then forms part of the lower wall of the TJE nozzle. At present, this short splitter plate is set at zero angle, i.e., horizontally. Beyond the end of this splitter plate, the flow expands into the free stream flow or, in case the DMSJ nozzle is also operational, into the plume from the DMSJ nozzle exit.

\begin{figure}[tb]
    \centering
    \includegraphics[width=0.5\linewidth]{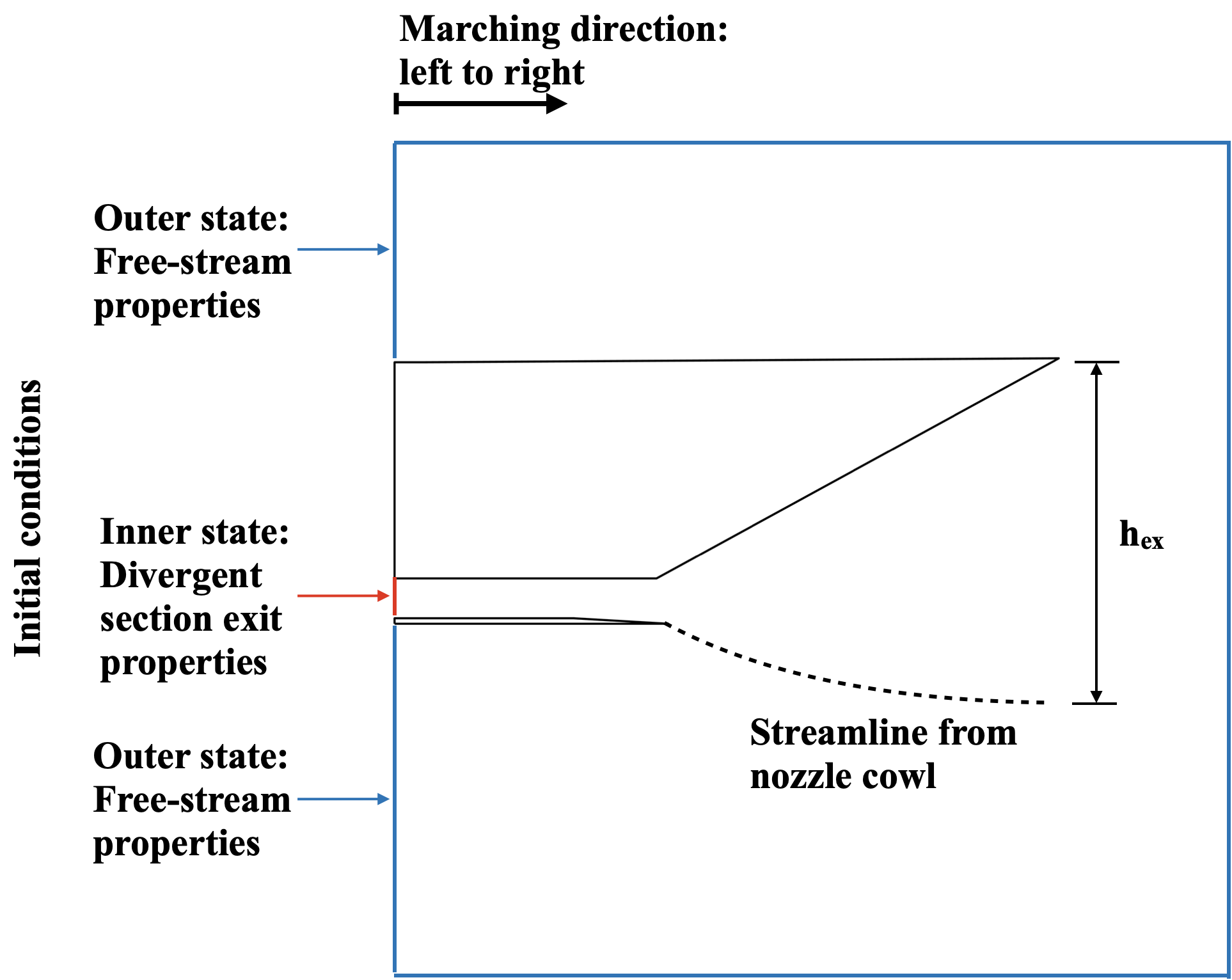}
    \caption{Specification of boundary conditions for nozzle flow domain for supersonic flow analysis.}
    \label{fig:ROAM_nozzle_domain}
\end{figure}

The flow in the circular-to-rectangular transition segment, which is downstream of the throat (station 6t) is approximated as isentropic. From there on, the supersonic flow downstream in the rest of the nozzle up to the exit of the SERN ramp is computed using a version of the method of characteristics, described below. The nozzle domain for supersonic flow analysis is shown in Fig.~\ref{fig:ROAM_nozzle_domain}.
 
\subsection{Reduced-Order Aerodynamic Modeling} \label{sec:ROAM}
Reduced-Order Aerodynamic Modeling (ROAM) is a method-of-characteristics-based inviscid code that is used for quick computation of supersonic flow in two-dimensional domains. It is implemented using the reduced-order methodology in Dalle et al.~\cite{2010Dalle-1}, where instead of discretizing the domain into smaller cells and performing numerical analysis using methods such as finite volume schemes, one directly propagates the upstream values along characteristic lines, at the same time solving a series of Riemann problems for interactions between shocks and expansion waves. The continuous Prandtl-Meyer expansion fan is modeled with a finite number of expansion waves. This approach is a generalized and automated version of the method of characteristics. Details of the procedure are given in Dalle et al.~\cite{2010Dalle-1} and are not repeated here. Our implementation of ROAM is first tested against an Euler CFD solver below for a very similar nozzle configuration as the one used in this study.

\begin{figure}[b!]
    \centering
    \includegraphics[width=0.45\linewidth]{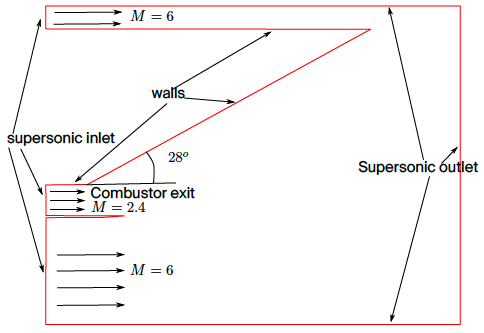}
    \caption{Test case nozzle flow and geometry used for validation of ROAM.}
    \label{fig:ROAM_nozzle_SU2_domain}
\end{figure}

\begin{figure}[tb]
     \centering
     \begin{subfigure}[b]{0.5\textwidth}
         \centering
         \includegraphics[width=\textwidth]{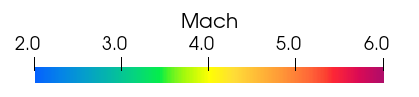}
     \end{subfigure}
     \\
     \begin{subfigure}[b]{0.4\textwidth}
         \centering
         \includegraphics[width=\textwidth]{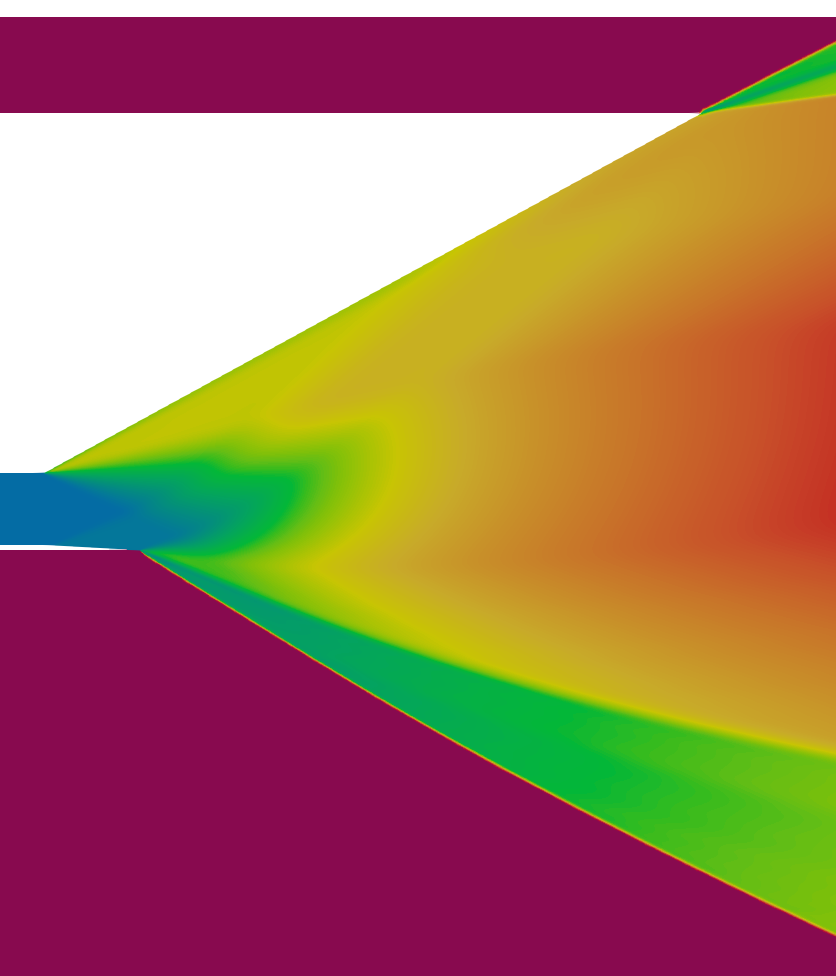}
         \caption{Inviscid SU2}
         \label{fig:MachNozzleSU2}
     \end{subfigure}
     \hfill
     \begin{subfigure}[b]{0.4\textwidth}
         \centering
         \includegraphics[width=\textwidth]{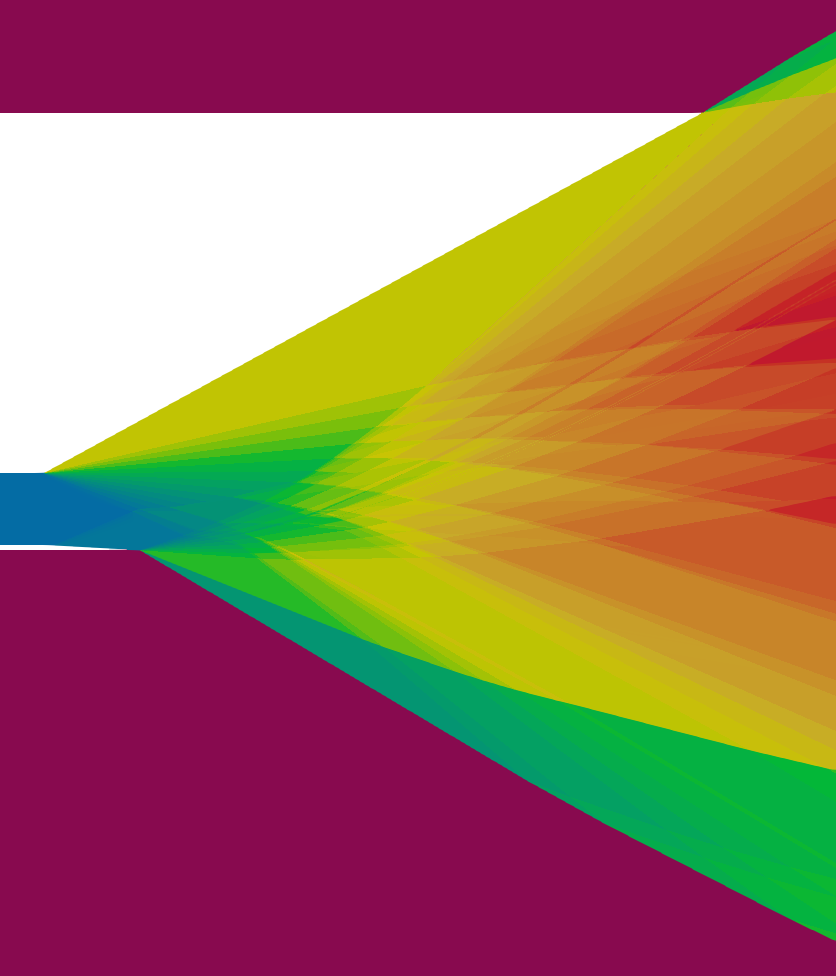}
         \caption{ROAM}
         \label{fig:MachNozzleROAM}
     \end{subfigure}
        \caption{Comparison of supersonic nozzle Mach contours obtained using (a) Inviscid SU2 computation, versus (b) ROAM solver.}
        \label{fig:MachNozzle}
\end{figure}

A hot, supersonic flow of Mach 2.4 from the combustor exit, experiencing a sudden expansion of $28$ deg through a nozzle, with free stream conditions of Mach 6, is used for validating the ROAM solver. An open-source CFD solver, namely the SU2 code (Economon et al.~\cite{economon2016su2}), is employed to simulate the inviscid flow field and serves as the reference for the validation. The computational domain and boundary conditions used for the inviscid computation are as shown in Fig.~\ref{fig:ROAM_nozzle_SU2_domain}.
Mach contours depicted in Fig.~\ref{fig:MachNozzle} show that an expansion fan forms at the expansion corner on the top surface. 
Another expansion fan occurs at the tip of the bottom cowl surface where a shear layer is shed downstream. This shear layer acts as a compression surface for the free stream flow under the cowl surface, hence a shock is formed. All these flow features are well predicted by the  ROAM solver, including the interaction between the expansion fans running between the top wall and the lower shear layer. For this analysis, the continuous Prandtl-Meyer expansion is discretized with an increment of 2~deg in ROAM, which explains the granular interactions seen in Fig.~\ref{fig:MachNozzleROAM}. The average flow properties calculated at the nozzle exit plane from the SU2 and ROAM simulations are compared in Table~\ref{tab:nozzle}. It is clear that ROAM predicts the flow parameters to an accuracy better than $5\%$ and the thrust is accurate to within $1\%$. 
\begin{table}[b]
    \centering
    \def\arraystretch{1.6}%
    \caption{Comparison of average nozzle exit properties from SU2 and ROAM solvers for the test case in Fig.~\ref{fig:ROAM_nozzle_SU2_domain}}
    \begin{tabular}{cccc}
    & SU2 (Inviscid) & ROAM solver & Relative error (\%) \\ \hline 
    $\rho_\infty/\rho_e$  & $10.952$ & $11.454$ & $4.58$ \\ 
    $p_\infty/p_e$ & $29.467$ & $28.712$ & $2.56$ \\ 
    $T_\infty/T_e$ &  $2.751$ & $2.708$ & $1.56$ \\ 
    $M_e$ & $4.947$ & $4.972$ & $0.51$ \\ 
    $\mathrm{Thrust}_{e}\ (kN)$ & $14.5$  & $14.4$ & $0.68$ \\ 
    \end{tabular}
    \label{tab:nozzle}
\end{table}

\subsection{Thrust Calculation}
To compute the thrust generated by the nozzle, the average properties at the nozzle exit station are extracted from the ROAM analysis. The nozzle exit plane is the vertical plane at the trailing edge of the TBCC geometry (refer Fig.~\ref{fig:ROAM_nozzle_domain}). Since the bottom surface ends much before the nozzle exit plane, the hot flow streamtube needs to be traced from the nozzle cowl, with the dashed streamline in Fig.~\ref{fig:ROAM_nozzle_domain} forming the lower (virtual) wall of the TJE nozzle. The height of the exit plane is determined based on this streamline and the trailing edge of the SERN ramp. Average properties are then calculated for this nozzle exit flow area. Using these average properties, the thrust is calculated as follows:

From steady flow analysis, the force acting on the exit plane is given by
\begin{equation} \label{thrust:fx}
    F_x = m_eu_e - m_au_a + (P_e-P_{\infty})A_{ex}
\end{equation}
\begin{equation} \label{thrust:fy}
    F_y = m_ev_e - m_av_a + (P_e-P_{\infty})A_{ey}
\end{equation}
where, $A_{ex} = h_{ex} \times w$, $A_{ey} = 0$, and area, mass flow rate, and axial, normal velocities are represented by $A$, $\dot m$, $u$ and $v$, respectively. The subscripts $a$, $e$, $x$ and $y$ represent air, exit, x-direction and y-direction, respectively. The axial component of force, $F_x$, is the thrust that is of interest to us. The normal force $F_y$ (and the resultant pitching moment) is assumed to be already included in the aerodata. The air mass flow rate $\dot m_{a}$ is determined from the known value at engine entry ($\dot m_c$), and the mass flow rate at the nozzle exit $\dot m_{e}$ is the value available at the turbine exit.

\subsection{Application to TJE nozzle}
 For the present analysis, the TJE nozzle flow is considered independently, without simulating the exhaust flow through the DMSJ nozzle simultaneously. Hence, the outer domain in Fig.~\ref{fig:ROAM_nozzle_domain} is initialized with free stream conditions. In practice, during the transition between turbojet and ram modes, both the DMSJ and TJE may be operational in a certain time window, hence the exhaust gas from each nozzle will interact and influence the thrust produced by the other (Mo et al.~\cite{mo2014design}).

\begin{figure}[tb]
    \centering
    \includegraphics[width=0.8\linewidth]{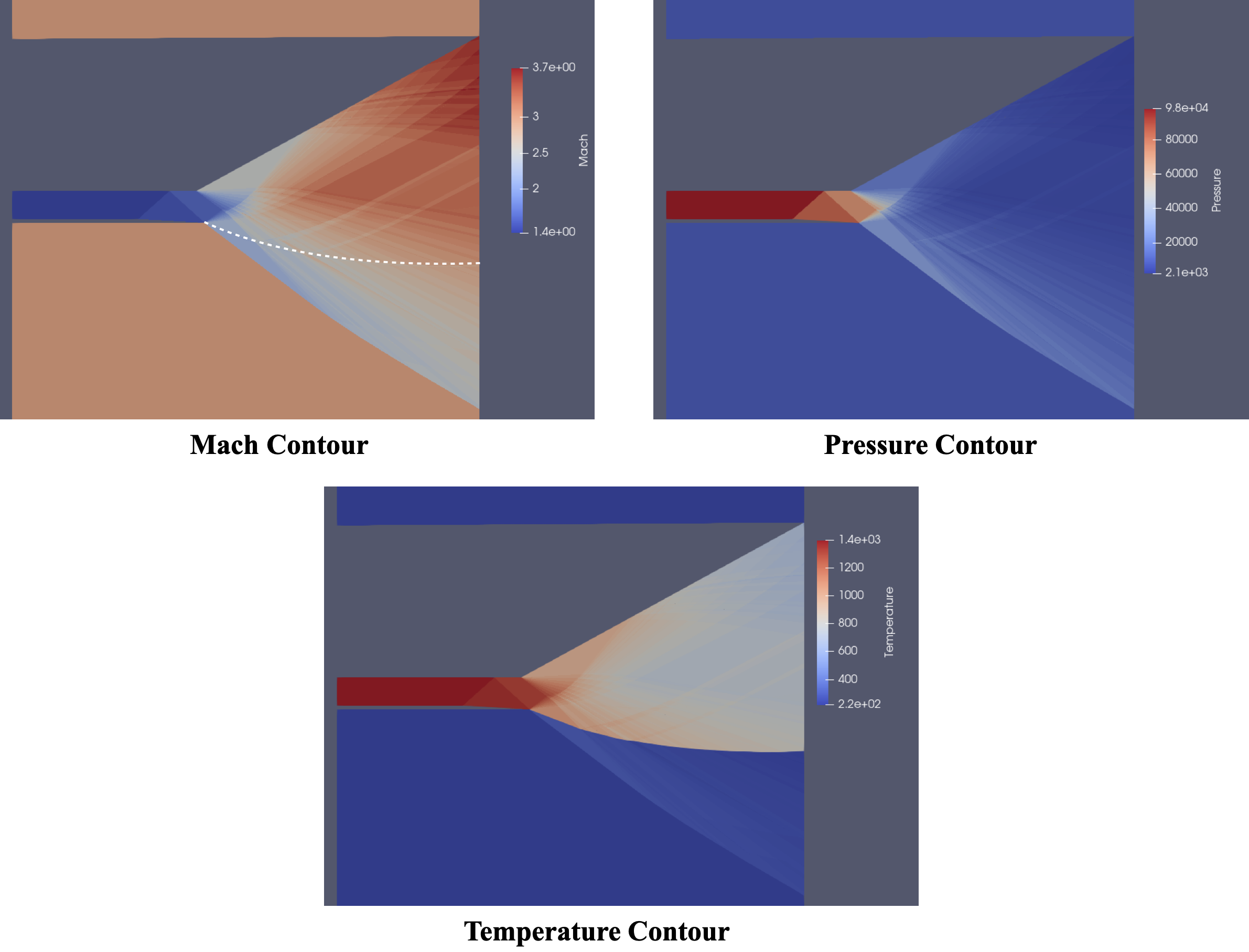}
    \caption{Nozzle flow field at $M_\infty$ = 3, AoA = 0$^{\circ}$, at 18.1 km altitude using the ROAM solver.
    }
    \label{fig:FlowField_Nozzle_M3}
\end{figure}

The ROAM code is now applied to the TJE nozzle geometry as shown in Fig.~\ref{fig:ROAM_nozzle_domain}. Contours of Mach number, pressure and temperature from the ROAM solution for the nozzle at TJE handover condition, $M_\infty = 3$, angle of attack 0 deg, at 18.1 km altitude, are presented in Fig.~\ref{fig:FlowField_Nozzle_M3}. 
The boundary between the hot TJE nozzle flow and the colder free stream flow is apparent in the temperature contour. This boundary is also marked by a white dotted line in the Mach number contour where one can discern the expansion fans in the hot nozzle flow reflecting off this dividing streamline. The free stream flow undergoes a compression when encountering this streamline causing the shock that is visible in Fig.~\ref{fig:FlowField_Nozzle_M3}. At the nozzle exit plane, the region between the tip of the SERN ramp and the end of the white dotted streamline forms the exit area.

\section{Intake Modeling and Simulation} \label{sec:Intake}
\begin{figure}[tb]
    \centering
    \includegraphics[width=0.8\linewidth]{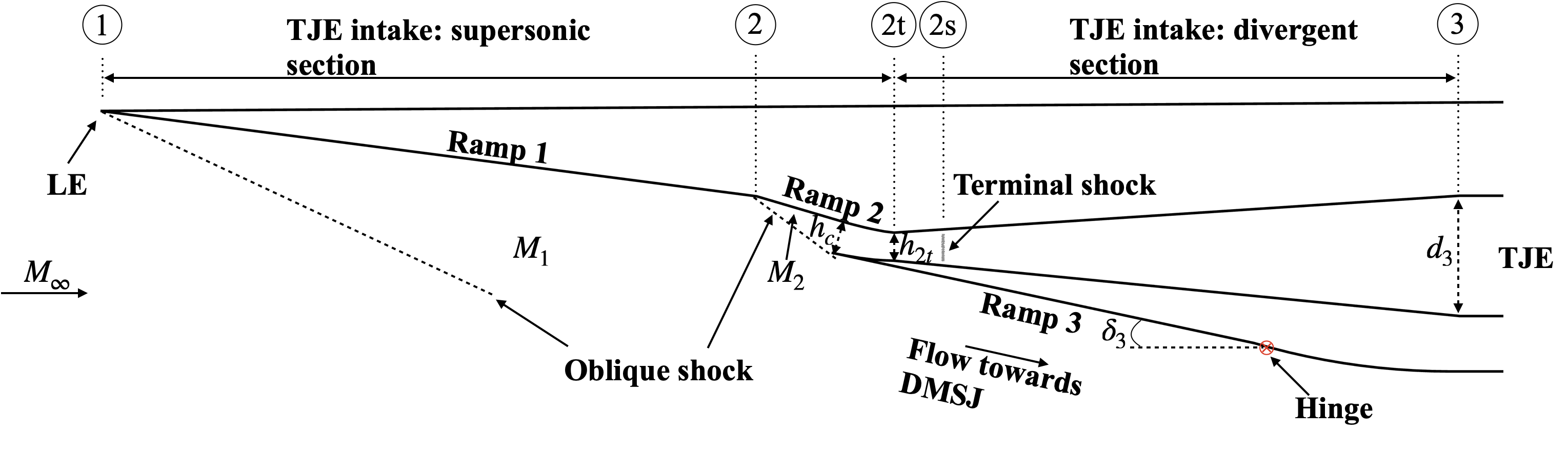}
    \caption{Arrangement of intake ramp, shocks, and splitter plate for turbojet intake of TBCC engine.}
    \label{fig:Intake_modeling}
\end{figure}

The design and operation of the supersonic intake during the turbojet mode of the TBCC is described in this section.
As noted in Fig.~1, the intake has been optimally designed with two ramps, with the ramp angles selected corresponding to Mach 6 cruise flight. Understandably, these ramp angles are much smaller than optimal for the case of Mach 3 flight, which is the point of interest for the TJE. For the first ramp angle of $7.4$ deg, the oblique shock angle at Mach 6 is $15.2$ deg, whereas at Mach 3 it is $25.1$ deg, as shown (slightly exaggerated) in Fig.~\ref{fig:Intake_modeling}. 
The flow deceleration is thus also less than desired with $M_1=2.63$. Ordinarily, in DMSJ mode, this flow would then pass over the second ramp of angle $16.5$ deg, slowing down further, with additional deceleration taking place internally within the intake duct. The second ramp would normally block the entry of the free stream flow to the TJE. When operating in turbojet mode, part of the second ramp is rotated about a hinge at its aft end, as marked in Fig.~\ref{fig:Intake_modeling}. This opens up a passage for the flow to enter into the TJE channel with the hinged segment acting as a splitter plate and the tip of this hinged segment being the cowl lip for the turbojet intake channel. The splitter plate is rotated to such an extent that the shock from the second ramp sits exactly at this cowl lip at free stream conditions of Mach 3, zero angle of attack. That is, there is no spillage of the flow at this cowl lip as far as the TJE is concerned. The rotation of the hinged segment creates a new ramp (called "Ramp 3" in Fig.~\ref{fig:Intake_modeling}), and the flow through the first ramp shock, followed by the third ramp shock, enters the DMSJ intake channel. This is of relevance when the TJE and DMSJ modules are operating simultaneously; however, we do not analyze this condition here.

The length of the splitter plate is determined by two factors: i) the required air mass flow rate to generate the necessary thrust, and ii) shock-on-cowl-lip condition. To see this, consider the splitter plate of arbitrary length as sketched in Fig.~\ref{fig:Intake_modeling}. With the second ramp shock hitting this splitter plate at its tip (cowl lip), the lip height, $h_c$, is determined, which dictates the mass flow rate ($\dot m_c$) entering the TJE intake duct. Note that the TJE intake flow is still supersonic, so a further converging section is required until a throat of height $h_{2t}$ is formed (marked in Fig.~\ref{fig:Intake_modeling}). Beyond this is a subsonic diffuser leading to the compressor face, and the terminal normal shock (station 2s) is usually located in this diffuser section, but as close to the throat as possible. If the TJE core is being sized for a higher mass flow rate, then the length of the splitter plate must be reduced such that, when it intercepts the second ramp shock at its tip, the height $h_c$ is larger, thus allowing a larger mass flow rate into the TJE intake duct. It is important to note that as the TJE core is resized (during the iterative solution procedure), the TJE intake also needs to be modified for each iteration to resize the splitter plate.

To ensure smooth operation of the mixed compression inlet and to prevent unstart problems, the throat area must be sized to swallow the normal shock by the intake effectively. The required height of the duct at the throat ($h_{2t}$) depends on the height at the cowl ($h_c$) and the required contraction ratio (CR) for the constant-width engine, where $CR=h_c/h_{2t}$. Since the present calculations are being done for only two values of the Mach number, 2.5 and 3, the shock swallowing requirement is being imposed at the lower Mach number of $2.5$. The CR is calculated using the modified Kantrowitz criterion as follows (Flock and G{\"u}lhan~\cite{flock2019modified}):
\begin{equation} \label{Eq:CR1}
    \frac{1}{CR}=SI\left(\frac{1}{CR_{K}}-\frac{1}{CR_{is}}\right)+\frac{1}{CR_{is}}
\end{equation}
where,
\begin{equation} \label{Eq:CR2}
    CR_{is} = \frac{A}{A^*} = \frac{1}{M} \left[ \frac{2}{\gamma+1} \left( 1+\frac{\gamma-1}{2}M^2 \right)  \right]^{\frac{\gamma+1}{2(\gamma-1)}}
\end{equation}
 and 
 \begin{equation} \label{Eq:CR3}
     CR_{K} =  \frac{A}{A^*} = \left[ \frac{(\gamma+1)M^2}{(\gamma-1)M^2+2} \right]^{0.5} \left[ \frac{(\gamma+1)M^2}{2 \gamma M^2-(\gamma-1)} \right]^{1/(\gamma-1)}
 \end{equation}
where $A^*$ is the intake throat area at the sonic condition, $SI=0.685$ is an empirical constant, and the subscripts $is$ and $K$ stand for `isentropic' and `Kantrowitz,' respectively.  For the given Mach number at the cowl lip ($M_2$), the minimum $h_{2t}$ can be calculated using Eqs.~\ref{Eq:CR1}-\ref{Eq:CR3}. For example, at free stream Mach number 2.5, the Mach number at the cowl lip is 1.852, and the $CR_{is}$ and the $CR_K$ are calculated to be 1.4976 and 1.182, respectively. Then, CR from the cowl to the throat is calculated from Eq.~\ref{Eq:CR1} to be 1.226.

\subsection{ROAM intake computation}
\begin{figure}[tb]
    \centering
    \includegraphics[width=0.8\linewidth]{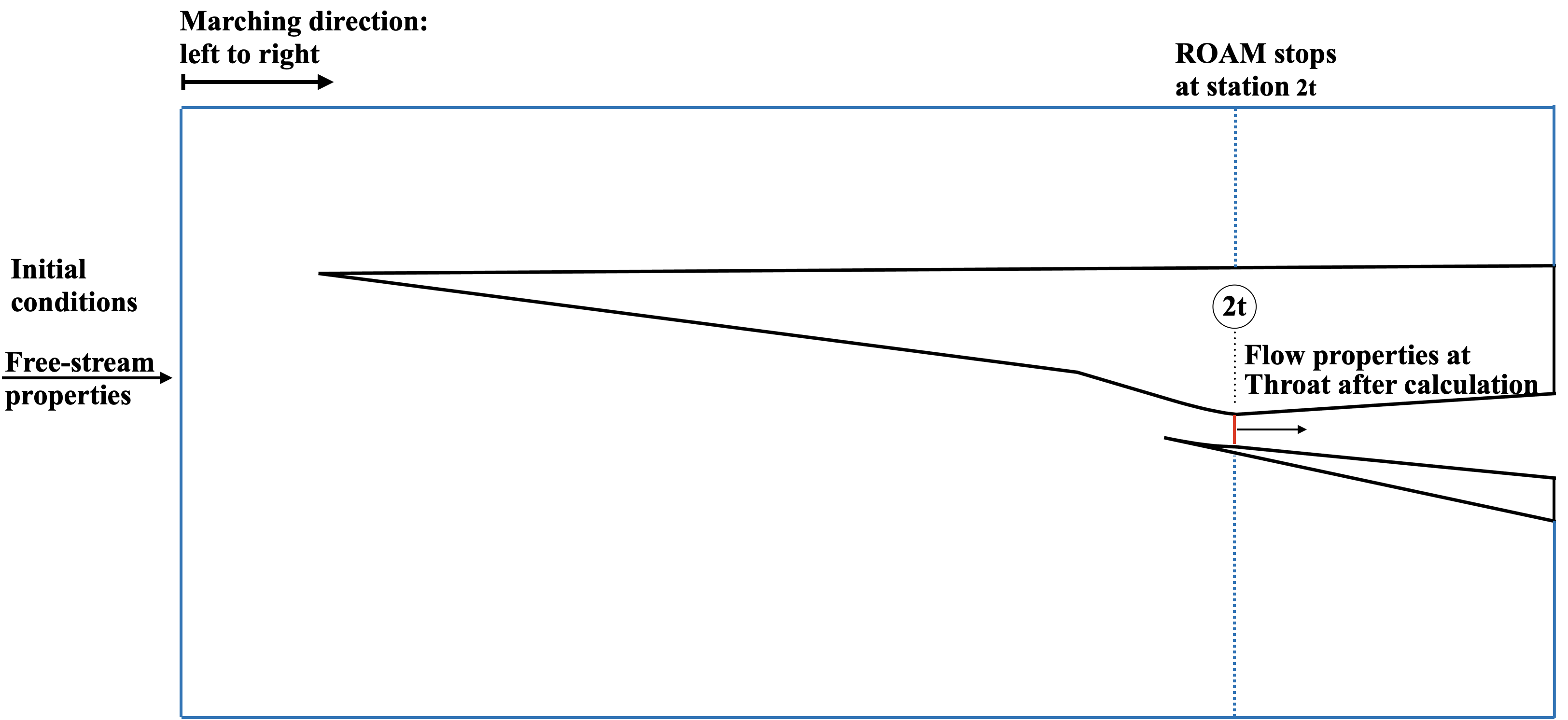}
    \caption{Specification of the intake flow domain for supersonic flow analysis.}
    \label{fig:ROAM_intake_domain}
\end{figure}

The same inviscid ROAM code, described in Sec.~\ref{sec:ROAM}, that was used for the nozzle flow computation, is also employed to calculate the intake flow properties. The flow domain for ROAM intake calculation is as shown in Fig.~\ref{fig:ROAM_intake_domain}. Starting with the free stream conditions, ROAM computes the supersonic flow over the intake ramps and over the cowl up to the throat section. 
The average properties at the throat are obtained for use as the inlet conditions for the subsonic diffuser (described below). 
 
{\bf A note on the selection of throat height, $h_{2t}$}: To begin with, for the sizing of the intake at each iteration, the required throat height $h_{2t}$ is calculated as per Eqs.~\ref{Eq:CR1}-\ref{Eq:CR3}. Then, the ROAM code is run for the corresponding intake geometry. However, in certain instances, the ROAM code does not compute the flow all the way to the throat and terminates upstream. While this may be due to several reasons, during the iterative procedure it is not desirable to pause and investigate the problem. Instead, it is assumed that this has happened because ROAM fails to find an acceptable supersonic solution at that point, and that this is mostly due to intake unstart, where the terminal normal shock has moved upstream from its regular position. The intake is then automatically resized to allow for a larger throat area until ROAM is again able to produce a solution under all given free stream conditions.
 
\begin{figure}[tb]
    \centering
    \includegraphics[width=0.8\linewidth]{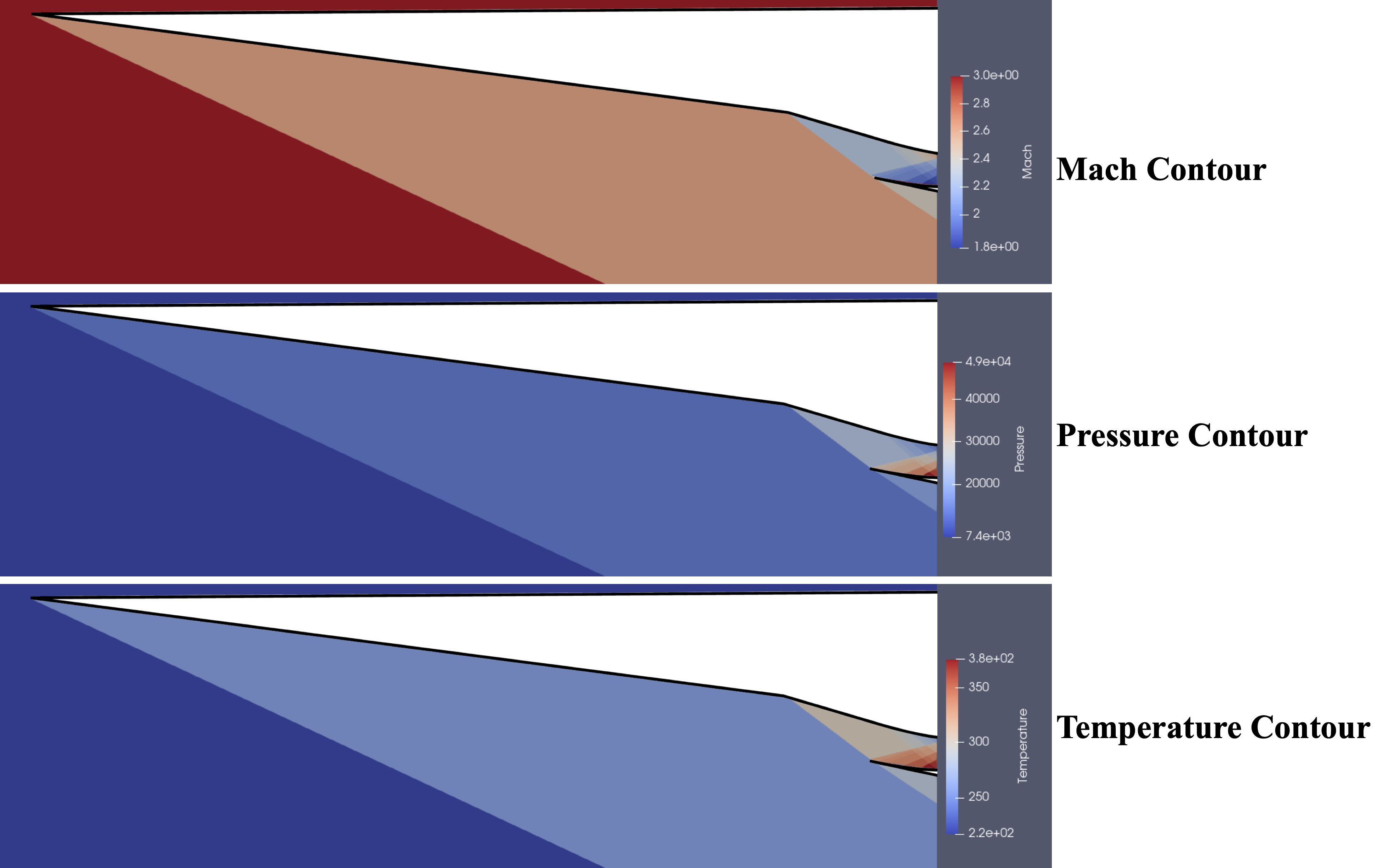}
    \caption{Intake flow field at $M_\infty$ = 3, AoA = 0$^{\circ}$, at 18.1 km altitude using the ROAM solver.}
    \label{fig:FlowField_Intake_M3}
\end{figure}

Typical ROAM results for the supersonic section of the intake are shown in Fig.~\ref{fig:FlowField_Intake_M3} which shows the compression of the incoming air by two external oblique shocks and the internal compression surface up to the throat. The Mach contour clearly reveals the oblique shock from the second ramp touching the cowl lip in this case (design condition). The ROAM solver stops just short of the intake throat section, where the flow is still supersonic in all cases.

\subsection{Intake subsonic diffuser}
The averaged supersonic flow properties at the throat station are available from the ROAM solution. Typically, one would expect the intake system to be operating either in critical mode or supercritical mode, depending on the back pressure at the intake exit. Consequently, the terminal shock either stands at the throat ($A_{2s}/A_{2t} = 1$ \textemdash critical) or, more often, between the throat and the diffuser exit ($1 < A_{2s}/A_{2t} \leq A_{3}/A_{2t}$ \textemdash supercritical) (see Fig.~12 where the Station `2s' refers to the location of the terminal normal shock, and Station `3' is the exit of the subsonic diffuser). The flow in the subsonic diffuser is taken to be isentropic except across the normal shock, of course. The terminal shock location, which satisfies the imposed back pressure, is determined by using the isentropic area-Mach relation:
\begin{equation} \label{Eq:A2byA1}
    \frac{A_j}{A_i} = \frac{M_i}{M_j} \left( \frac{1+\frac{\gamma-1}{2}M_i^2}{1+\frac{\gamma-1}{2}M_j^2}\right)^{-\frac{\gamma+1}{2(\gamma-1)}}
\end{equation}
where the subscripts `i' and `j' represent the inlet section where properties are available and the outlet section where the properties are being calculated, respectively. Taking the area ratio between the normal shock station as outlet and the throat as inlet (that is, $(A_j/A_i) = (A_{2s}/A_{2t})$), the supersonic Mach number ($M'_{2s}$) upstream of the normal shock is calculated by solving Eq.~\ref{Eq:A2byA1} using the bisection root-finding method. The other flow properties ahead of the terminal shock are calculated using the isentropic relations:
\begin{equation} \label{Eq:isen1}
    \frac{P'_{j}}{P_{0i}} = \left(1 + \frac{\gamma - 1}{2} {M'_{j}}^2\right)^{-\frac{\gamma}{\gamma - 1}}
\end{equation}
\begin{equation} \label{Eq:isen2}
    \frac{T'_{j}}{T_{0i}} = \left(1 + \frac{\gamma - 1}{2} {M'_{j}}^2\right)^{-1}
\end{equation}
The flow properties behind the terminal shock are then obtained using the normal shock relations:
\begin{equation} \label{Eq:norm1}
    M''_{j} = \sqrt{\frac{(\gamma-1){M'_{i}} ^2 + 2}{2 \gamma {M'_{i}}^2-(\gamma-1)}}
\end{equation}
\begin{equation} \label{Eq:norm2}
    P''_{j} = P'_{i}*{\frac{2 \gamma {M'_{i}}^2-(\gamma-1)}{\gamma+1}}
\end{equation}
\begin{equation} \label{Eq:norm3}
    T''_{j} = T'_{i}\ *\ \frac{[(\gamma-1){M'_{i}}^2 + 2] * [2 \gamma {M'_{i}}^2-(\gamma-1)]}{[(\gamma+1)M'_{i}]^2}
\end{equation}
The conditions immediately ahead of and behind the terminal shock are represented by superscript $'$ and $''$, respectively. Then defining the station downstream of the normal shock as the inlet `i' and the subsonic diffuser exit as the outlet `j', the diffuser exit flow (Station 3) properties are calculated using Eq.~\ref{Eq:A2byA1} with $A_j/A_i = A_3/A_{2s}$, and the other properties follow by use of Eqs.~\ref{Eq:isen1} and \ref{Eq:isen2} appropriately. In addition to the loss across the normal shock, a $5\%$ total pressure loss is considered to account for viscous losses and the loss due to intake cross-section change from rectangular at the throat to circular at the compressor face (Trefny and Benson~\cite{trefny1995integration}). Typical intake flow properties at free stream conditions of $M_\infty = 3$, AoA = 0~$\deg$, at 18.1 km altitude, with the terminal shock close to its critical position, are tabulated in Table~\ref{tab:Stn_properties} at various stations.
\begin{table}[b]
    \centering
    \caption{Typical intake flow properties under free stream conditions of $M_\infty = 3$, AoA = 0~$\deg$, at 18.1 km altitude}
    \label{tab:Stn_properties}
    \begin{tabular}{llll}
        \textbf{Station} & \textbf{$P$}$\ (Pa)$ & \textbf{$T$}$\ (K)$ & \textbf{$M$} \\ \hline
         Throat ($2t$) & $31187.1$ & $333.4$ &$ 2.035$ \\ 
        Behind terminal shock ($2s$) & $145512.3$ & $572.3$ & $0.572$ \\ 
        Subsonic diffuser exit ($3$) & $173895.7$ & $602.2$ & $0.25$  \\ 
    \end{tabular}
\end{table}
These numbers are re-computed at every iteration of the turbojet engine sizing process with the intake suitably modified as described earlier.

\section{Turbojet Module Sizing} \label{sec:TJE_sizing}
With the design and modeling of the individual components (intake, engine core, and nozzle) complete, the next step is to combine the component models into a single integrated model to simulate the flow through the turbojet module and to evaluate the performance of the TJE. However, the component models are inextricably interconnected in various ways making the integrated model fairly complex in terms of the physics. For the given free stream conditions from the flight trajectory, the thrust generated by the engine is calculated by adjusting the fuel flow rate to the turbojet engine. The subsonic diffuser section of the intake and the engine core module work together to determine the location of the terminal shock for the given fuel flow rate (or equivalence ratio) while maintaining the constraint on the TIT. The objective of the TJE sizing is to obtain the desired thrust at the handover condition to allow for a smooth mode transition between the turbojet and ramjet modes, as suggested by Kashif and Varnavas~\cite{javaid2005thrust}. The transition from turbo to ram mode is planned for Mach 3 and an angle of attack of zero while maintaining constant thrust. Therefore, the TJE must be sized to produce a thrust of 1568~N, with a margin of error of $5\%$ to allow for the various inaccuracies in the modeling process. In the following, the sizing calculations are carried out by simulating the engine core concurrently with the intake and nozzle, in an iterative manner until a value of thrust within $5\%$ of the target thrust is achieved without violating any of the constraints. 

Before carrying out the sizing process, certain parameters are chosen and fixed throughout. At the handover point, the intake is considered to be started and operating in critical mode, which means that the terminal shock is located at the throat (i.e., $A_{2s}/A_{2t} = 1$). The hinge point of the intake splitter plate is positioned 1539.04 mm axially aft of and 310.33 mm vertically downward from the vehicle nose (LE, as marked in Fig.~\ref{fig:Intake_modeling}). The intake duct segment between the cowl lip and the throat is shaped according to Connors and Meyers~\cite{Connors1956}, given the Mach number at the cowl lip ($M_2$). At handover, with free stream Mach number of 3, $M_2$ is calculated to be 2.25. The contraction ratio for the throat is calculated using Eqs.~\ref{Eq:CR1}-\ref{Eq:CR3} for the lowest free stream Mach number considered in this paper, which is Mach 2.5. This value of $CR$ represents the minimum required for the intake to start by swallowing the terminal shock. Following this ratio, the areas at the cowl lip and the throat are selected for the required mass flow rate at the handover Mach number of 3. Under this condition, the expected flow Mach number at the intake exit station is estimated to be 0.25, following a generic engine airflow schedule from Sanders and Weir~\cite{Sander2008}, and the exit duct area is accordingly calculated.

The parameters of the turbojet engine that remain unchanged from the baseline design include the compressor bleed air fraction, burner efficiency and pressure loss, turbine cooling flow, and duct pressure loss between the turbine and nozzle. An initial operating point is selected from the operating line previously calculated for the baseline engine in high-speed operation in Fig.~\ref{fig:J85_offdesign_compressor_map}, subject to the compressor exit temperature limit. At a free stream Mach number of 3, the air temperature at the compressor inlet reaches approximately 600~K. The compressor design pressure ratio (PR) is chosen to provide sufficient margin for adding heat in the combustion chamber, with the compressor exit temperature limited to about 1000 K. The acceptable pressure ratio is determined to be 5.025. The corresponding steady state parameters for the compressor and turbine are listed in Table~\ref{tab:Sizing_turbojet_design_parameters}.  
\begin{table}[b]
    \centering
    \caption{Initial choice of scaled operating parameters for turbojet engine}
    \label{tab:Sizing_turbojet_design_parameters}
    \begin{tabular}{ll|ll}
        \multicolumn{2}{c}{\textbf{Compressor}} & \multicolumn{2}{c}{\textbf{Turbine}} \\ \hline
        $N_c$ & $0.8846$ & $N_c$ & $100.54$ \\ 
        $PR$ & $5.025$ & $PR$ & $2.5911$ \\ 
        \it{Eff} & $0.8383$ & \it{Eff} & $0.8623$ \\ 
        \it{Rline} & $1.7437$ & $N$$\ (rpm)$ & $16540$ \\ 
    \end{tabular}
\end{table}

The maximum turbine inlet temperature is considered to be 2100 K. Jet fuel JP-10 is used, with a lower heating value of 43 MJ/kg. Gas properties such as the gas constant (R) and specific heat ratio ($\gamma$), necessary for further analysis, are computed using the online NASA CEA tool\footnote{https://cearun.grc.nasa.gov/}. Equivalence ratios and the combustor entry pressures are seen to vary from 0.45 to 0.58, and from 799792 to 1020425 Pa, respectively. Average values of the equivalence ratio (0.515) and combustor entry pressure (910108 Pa) are used to calculate R and $\gamma$ values from CEA, giving $R=287.7$~J/kg K and $\gamma=1.2553$. It is assumed that both R and $\gamma$ remain constant throughout the turbine and nozzle components. The convergence criterion is set to the absolute value of 0.005 for the iterative solver in T-MATS. 

\begin{figure}[tb]
    \centering
    \includegraphics[width=0.8\linewidth]{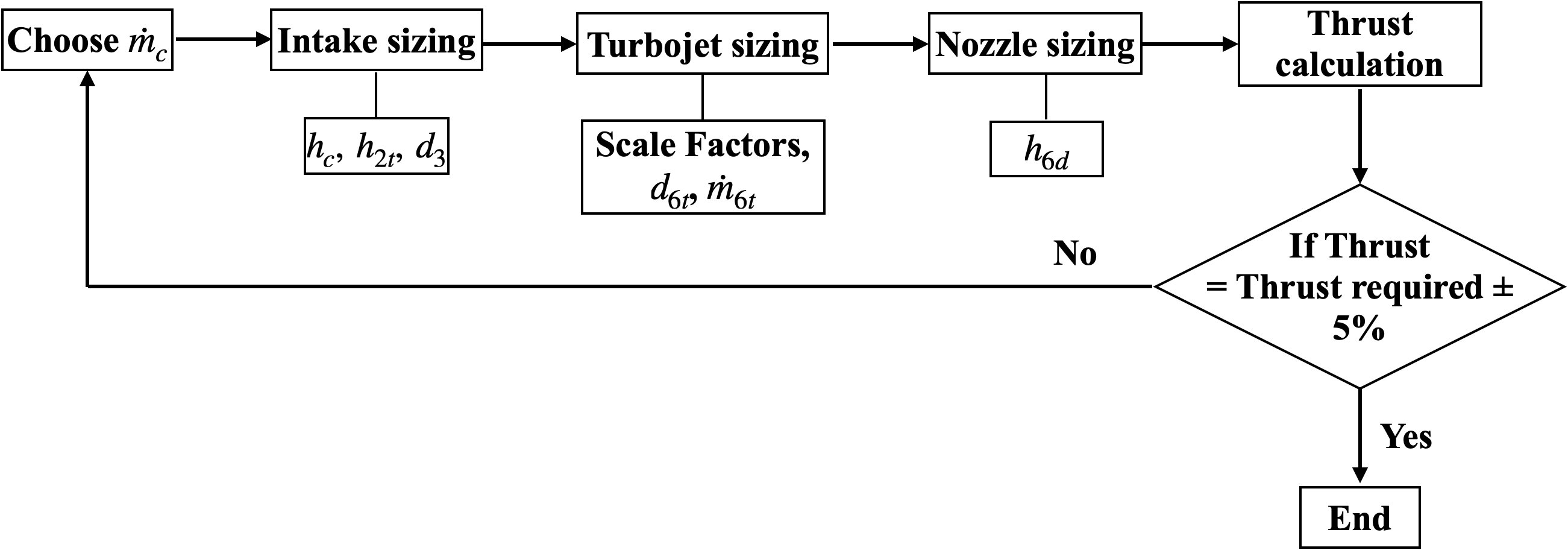}
    \caption{Flowchart for the turbojet engine iterative sizing procedure.}
    \label{fig:FlowChart_LSP_sizing}
\end{figure}

In order to size the turbojet engine for the desired thrust and the corresponding value of mass flow rate, the integrated model needs to be run iteratively as depicted in the flowchart of Fig.~\ref{fig:FlowChart_LSP_sizing} and explained below.
The sequence of steps to be followed is as below:
\begin{itemize}
    \item \textbf{Step 1}: Choose an air mass flow rate ($\dot m_{c}$) through the TJE flow path.
    \item \textbf{Step 2}: Size the intake for $h_c$, $h_{2t}$, and $d_3$, assuming the terminal shock to be located at the throat section. Check $h_{2t}$ for all operating conditions. If the ROAM solver terminates upstream of the throat, increase $h_{2t}$ until a ROAM solution is obtained for all operating conditions.
    \item \textbf{Step 3}: Resize the turbojet core for the chosen value of air mass flow rate ($\dot m_{c}$) by enabling {\it iDesign} for the scale factor of the performance maps. Adjust $\dot m_{f}$ and $\dot m_{6t}$ with the constraint that the TIT does not exceed 2100~K, and ensure the entry mass flow rate of turbojet at compressor entry equals the intake exit $\dot m_{3}$ (within the absolute difference of 0.001).
    \item \textbf{Step 4}: Size the nozzle for the divergent area exit height ($h_{6d}$). Use ROAM to get the flow field from the divergent duct exit to the nozzle exit.
    \item \textbf{Step 5}: Calculate thrust using the average properties from the nozzle exit.
    \item \textbf{Step 6}: Compare the calculated thrust with the required thrust. If the calculated thrust is not within the required limits, repeat the process from Step 1 to Step 6. Adjust $\dot m_{c}$ to match the required thrust\textemdash decrease $\dot m_{c}$ if the calculated thrust exceeds the requirement, or increase it if it falls short.
    
\end{itemize}

\begin{figure}[b]
    \centering
    \includegraphics[width=1\linewidth]{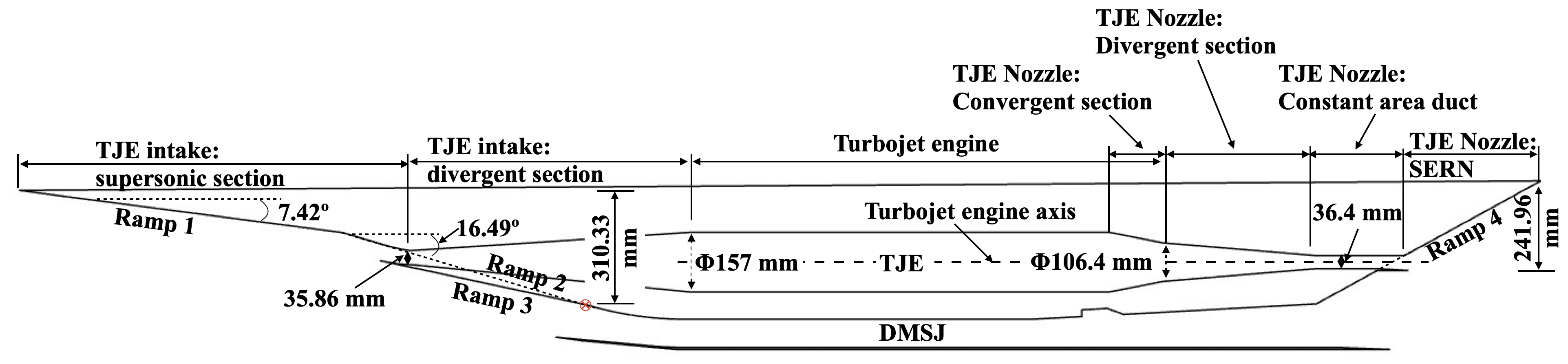}
    \caption{Final sized turbojet engine dimensions and integration with DMSJ module.}
    \label{fig:DMSJ_intake_with_dimensions}
\end{figure}

The mass flow rate and calculated thrust, along with parameters such as $h_c$, $h_{2t}$, $d_3$, $\dot m_{6t}$, $d_{6t}$, and $h_{6d}$, are tabulated in Table~\ref{tab:LSP_Sizing} for several iterations. 
\begin{table}[b!]
    \centering
    \caption{Converged solution from turbojet engine sizing iterations}
    \label{tab:LSP_Sizing}
    \begin{tabular}{lllll}
         & Iteration 1 & Iteration 2 & Iteration 3 & Iteration 4 \\ \hline
       $\dot m_c,\ kg/s$ & $3.247$ & $2.935$ & $2.725$ & $2.428$ \\
       $h_c,\ mm$ & $55.3$ & $50.1$ & $46.5$ & $41.2$ \\
       $h_{2t},\ mm$ & $46.8$ & $42.0$ & $39.75$ & $35.86$ \\
       $d_3,\ mm$ & $179.6$ & $170.1$ & $165.6$ & $157$ \\
       $\dot m_{6t},\ kg/s$ &$ 3.249$ & $2.936$ & $2.726$ & $2.429$ \\
       $d_{6t},\ mm$ & $121.6$ & $115.6$ & $111.9$ & $106.4$ \\
       $h_{6d},\ mm$ & $47.6$ & $43$ & $40.3$ & $36.4$ \\
       $Thrust,\ N $& $2309.7$ & $2062.9$ & $1891.3$ & $1640.7$ \\
       $Mismatch,\ \%$ & $47.3$ &  $31.6$ & $20.6$ & $4.6$ \\ 
    \end{tabular}
\end{table}
At the initial iteration with air mass flow rate of 3.247~kg/s, the thrust is 47.3\% too high. Further iterations reduce the mass flow rate and the mismatch in thrust until a solution is obtained with mass flow rate of 2.428~kg/s for which the thrust is within the 5\% margin of error. This last iteration fulfills our requirement, and the final scaled version of the turbojet engine is presented in Fig.~\ref{fig:DMSJ_intake_with_dimensions}. 
The scale factor for the performance maps for the final sized design are tabulated in Table~\ref{tab:TJE_SF}.
\begin{table}[tb]
    \centering
    \caption{Scale factor used for the performance maps for the final scaled turbojet engine} 
    \label{tab:TJE_SF}
    \begin{tabular}{lll}
        Scale Factor & \textbf{Compressor} & \textbf{Turbine}\\ \hline
        $N_c$ & $12852.790$ & $2.6758$ \\ 
        $W_c$ & $0.1311$ &$ 0.2016$ \\ 
        $PR$ & $1$ & $0.9154$\\ 
        \it{Eff} &	$0.9464$ & $1$\\  
    \end{tabular}
\end{table}
The final intake cowl and throat heights are 41.2~mm and 35.86~mm, respectively, and the intake exit diameter comes out to 157~mm for a Mach number of 0.25 at that station. With this diameter, the engine core is seen in Fig.~\ref{fig:DMSJ_intake_with_dimensions} to fit comfortably within the available volume above the DMSJ module. The TIT is 2093.5~K, marginally lower than the limit of 2100~K set. The nozzle throat diameter is calculated to be 106.4 mm, with the exit of the circular-to-rectangular divergent section having a height of 36.4~mm, followed by a constant-area duct of the same height extending to the start of the SERN ramp surface. The nozzle splitter plate is 241.96 mm below the trailing edge (TE) (as marked in Fig. \ref{fig:Nozzle_modeling}) in the final configuration.

\subsection{Off-Design Performance of TJE}
After sizing the TJE for the required thrust, the design parameters are frozen. The {\it iDesign} option in T-MATS is disabled, and the scale factors for the performance maps are updated. Off-design performance analysis is carried out at a free stream Mach of 2.5 with angle of attack ranging from $-2$ to $5$ deg, with Mach 3, angle of attack zero, representing the design condition. The intake exit properties at the critical mode of intake operation are provided in Table~\ref{tab:TJE_prop_stns} for all these cases. 
\begin{table}[b]
    \centering
    \caption{Flow properties at subsonic diffuser exit station with intake in critical mode of operation}
    \label{tab:TJE_prop_stns}
    \begin{tabular}{llllll}
        Mach &	AoA, deg &	$\dot m_{3}$, kg/s &	$P_{03}$, Pa  &	$T_{03}$, K &	$M_3$ \\ \hline
        $2.5$ & $-2$ & $2.411$ & $142694.84$ & $488.19$ & $0.268$ \\ 
        $2.5 $& $0$ & $2.558$ & $148012.15$ & $487.89$ & $0.273$ \\ 
        $2.5$ & $2$ & $2.676$ & $151786.47$ & $487.74$ & $0.279$ \\ 
        $2.5$ & $5$ & $2.840$ & $155684.67$ & $487.67$ & $0.290$ \\ 
        $3$ & $0$ & $2.428$ & $172513.26$ & $609.72$ & $0.25$ \\ 
    \end{tabular}
\end{table}
The mass flow rate of air through the TJE is calculated from the simulation of the supersonic part of the intake using ROAM. The mass flow rates of air at Mach 2.5 and Mach 3 at an angle of attack of zero are approximately close to each other because the density at Mach 2.5 is higher than at Mach 3 due to the difference in operating altitudes.

\begin{figure}[tb]
    \centering
    \includegraphics[width=0.5\linewidth]{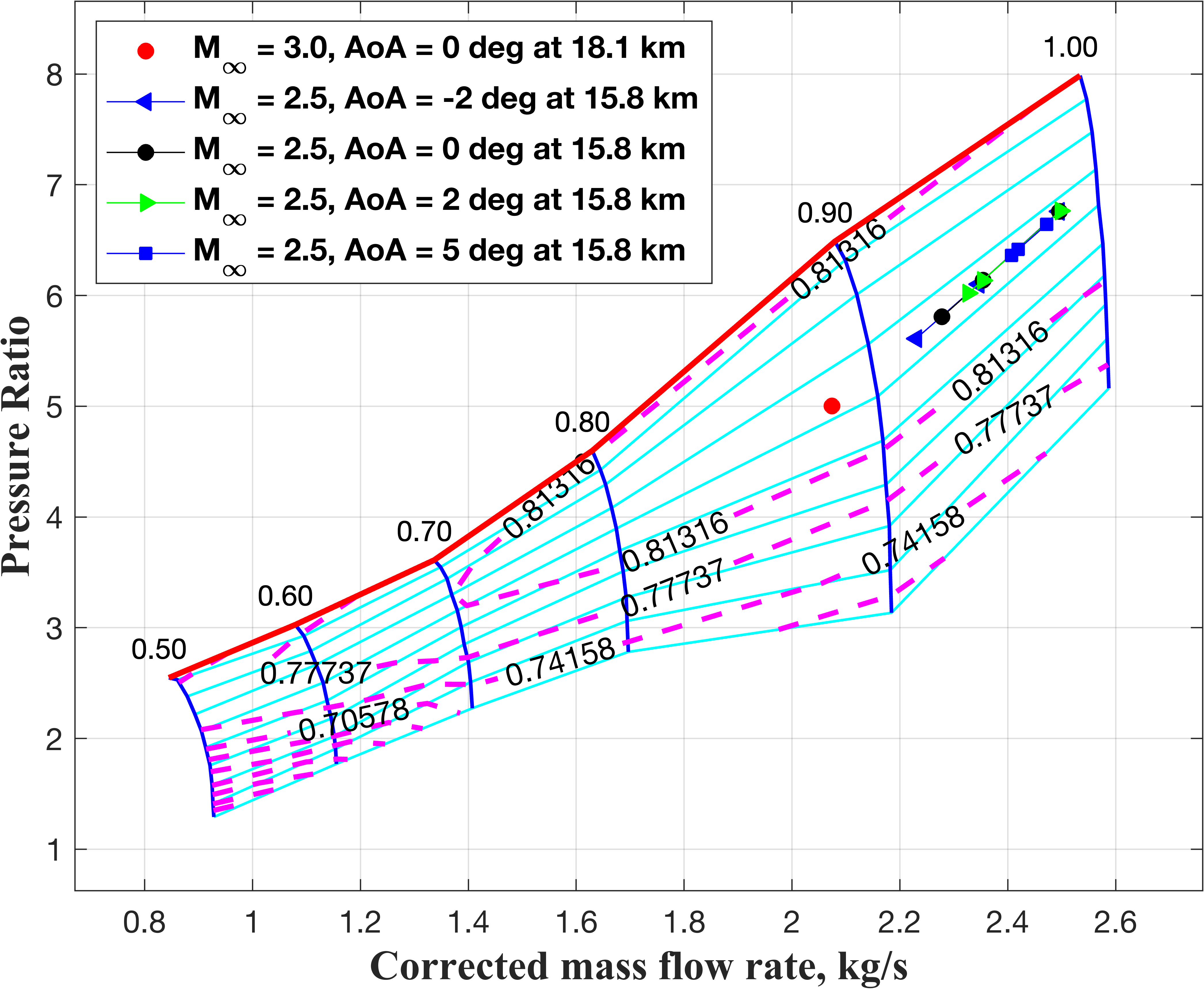}
    \caption{Compressor performance map at TJE design and off-design conditions.}
    \label{fig:TJE_comp_map_M2.5&3}
\end{figure}

The operating lines on the compressor map, constructed for Mach 2.5 and 3, are depicted in Fig.~\ref{fig:TJE_comp_map_M2.5&3}. 
For Mach 2.5, multiple steady states are found for each value of angle of attack. These are obtained by adjusting the fuel flow rate until the maximum TIT limit is hit, while the air mass flow rate is constant for a given Mach, Altitude, Angle of Attack combination. The design point at Mach 3 is also marked in Fig.~\ref{fig:TJE_comp_map_M2.5&3}. Interestingly, at this point the TIT is already close to its limiting value, hence increasing the fuel flow rate does not yield another steady state point. In fact, reducing the fuel flow rate also does not provide any additional steady states. When the fuel flow rate is reduced, it results in a discrepancy in the air mass flow rate between the intake duct and the turbojet, surpassing the maximum allowable error stipulated in Step~3 of the sizing process. Specifically, as the fuel flow rate decreases, the air mass flow rate demanded by the turbojet core also reduces, which can be managed only by flow spillage at the cowl lip or by bleeding a part of the air from the intake duct. Multiple operating points were indeed found on the compressor map in  Fig.~\ref{fig:J85_offdesign_compressor_map} for the original baseline engine, but that used a different, artificial intake which was not constrained by the design choices made for the DMSJ channel.

\begin{figure}[tb]
    \centering
    \includegraphics[width=0.5\linewidth]{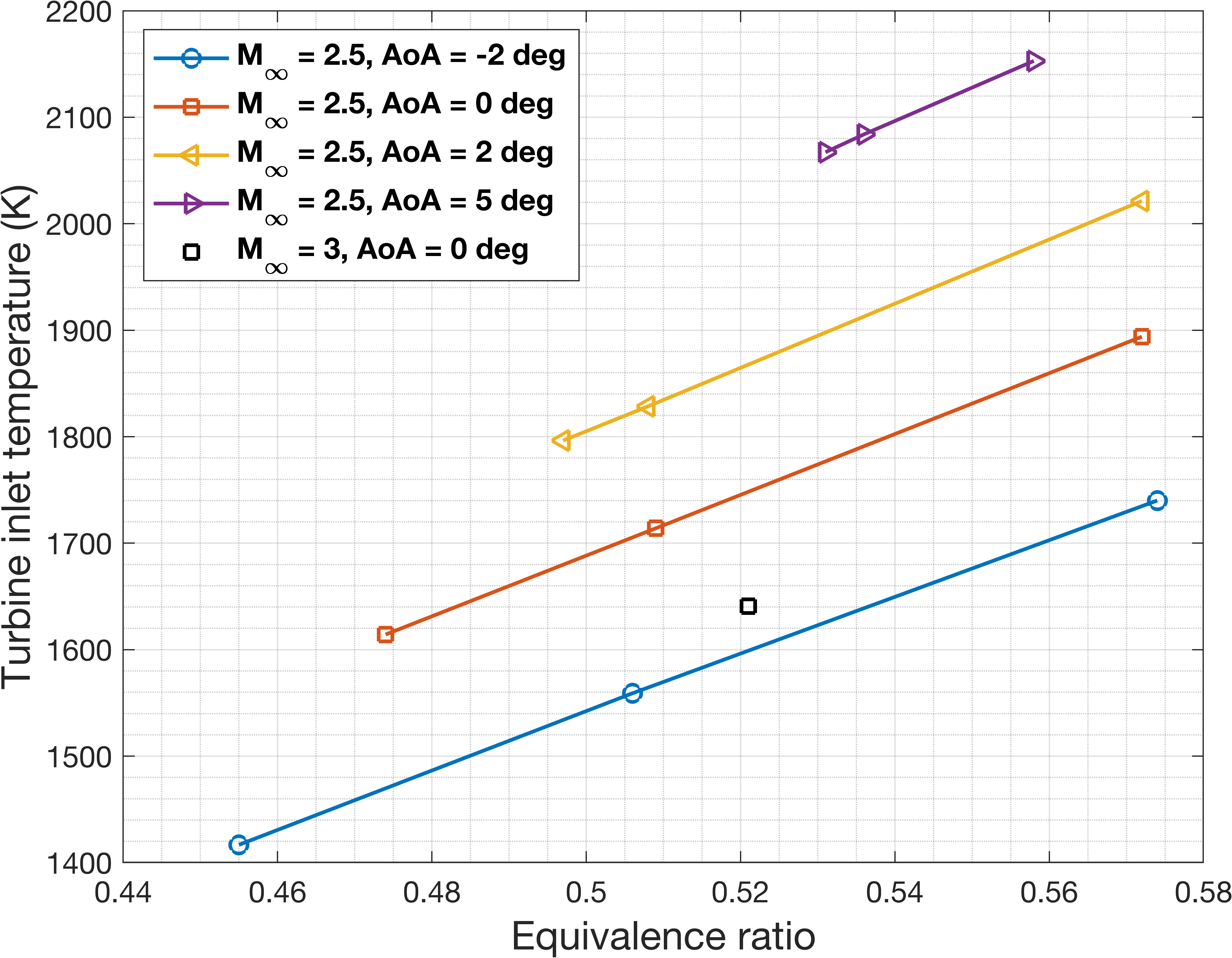}
    \caption{TJE thrust under off-design conditions (Mach 2.5 and various angles of attack); the design point Mach 3 thrust is also marked.}
    \label{fig:TJE_thrust}
\end{figure}

The thrust calculated at these design and off-design conditions is plotted in Fig.~\ref{fig:TJE_thrust} with varying fuel flow rate using the equivalence ratio on the X-axis.
Clearly, larger equivalence ratios yield higher thrust, until the limiting value of TIT is reached. The increased thrust for higher angles of attack is due to the larger mass flow rate into the intake with angle of attack. In this instance, fortuituously, a single steady state point was obtained for the Mach 3 condition with a value of thrust that also happened to be within the error margin of the ramjet thrust at handover. Under less fortunate circumstances, if no operating point could be located, or the thrust value exceeded the acceptable margin, then one would probably have to revise the handover strategy to a lower Mach number. Thus, as pointed out earlier in this article, the TJE sizing for integration into a TBCC engine is a fairly complex endeavor.

\section{Conclusion}
Starting from a given design of the scramjet part of a turbine-based combined cycle (TBCC) engine, this work has investigated the sizing of the turbojet module that satisfies the constraints imposed by the scramjet design and yet provides the required thrust at the turbo-ram handover point. A baseline turbojet engine has been modeled in T-MATS (Toolbox for the Modeling and Analysis of Thermodynamic Systems) and scaling factors for different engine sizes and mass flow rates worked out using the {\it iDesign} option in T-MATS. The intake and nozzle sections have been designed and linked with the engine core into a single integrated model. A Reduced-Order Aerodynamic Modeling (ROAM) solver is used to simulate the supersonic parts of the intake and nozzle flows. An iterative scheme is devised to size the TJE module such that the turbojet thrust at the handover Mach, Altitude closely matches the ram mode thrust of the DMSJ at that point while satisfying all the constraints. For the example considered in this work, a converged solution is indeed obtained with a value of thrust within an acceptable error margin of the desired handover thrust. However, in general, due to the various constraints, and given the Mach number and altitude of operation, it is quite likely that the TJE may not have a steady operating point at all, and even if it does, the intake may not be able to supply the required air mass flow rate, and even if that is possible, the nozzle expansion may not yield the desired thrust. Hence, the sizing of the turbojet module for a TBCC engine requires a degree of attention and finesse.

\begin{acknowledgement}
 We would like to thank Dr.~J.P.S.~Sandhu for providing access to the ROAM (Reduced-Order Aerodynamic Modeling) code for the intake and nozzle computational analysis.
\end{acknowledgement}

\bibliographystyle{siam}
\bibliography{cite}

\end{document}